\documentclass[manuscript]{aastex}
\usepackage{psfig}
\usepackage{rotating}
\usepackage{epsf}
\usepackage{amsmath}
\usepackage{textcomp}
\usepackage{color}
\usepackage[colorlinks=true,citecolor=blue]{hyperref}
\usepackage{graphicx,color}
\usepackage{times}
\usepackage{amssymb}

\slugcomment{Submitted to ApJ}

\shorttitle{1H 0323+342}
\shortauthors{Paliya et al.}

\begin{document}


\title{The Peculiar Radio-Loud Narrow Line Seyfert 1 Galaxy 1H 0323+342}


\author{Vaidehi S. Paliya\altaffilmark{1,2}, S. Sahayanathan\altaffilmark{3}, M. L. Parker\altaffilmark{4}, A. C. Fabian\altaffilmark{4}, C. S. Stalin\altaffilmark{1}, Ayesha Anjum\altaffilmark{5}, S. B. Pandey\altaffilmark{6}}





\altaffiltext{1}{Indian Institute of Astrophysics, Block-II, Koramangala, Bangalore-560034, India}
\altaffiltext{2}{Department of Physics, University of Calicut, Malappuram-673635, India}
\altaffiltext{3}{Astrophysical Science Division, Bhabha Atomic Research Center, Mumbai-400085, India}
\altaffiltext{4}{Institute of Astronomy, Madingley Road, Cambridge CB3 0HA, UK}
\altaffiltext{5}{Department of Physics, Christ University, Bangalore-560029, India}
\altaffiltext{6}{Aryabhatta Research Institute of Observational Sciences, Manora peak, Nainital-263129, India}



\begin{abstract}
We present a  multi-wavelength study of the radio-loud narrow line Seyfert 1 galaxy (NLSy1), 1H 0323+342, detected by {\it Fermi Gamma Ray Space Telescope}. Multi-band light curves show many orphan X-ray and optical flares having no corresponding $\gamma$-ray counterparts.  Such anomalous variability behavior can be due to different locations of the emission 
region from the central source. During a large flare, $\gamma$-ray flux 
doubling time scale as small as $\sim$ 3 hours is noticed. We built spectral energy distribution (SED) during different 
activity states and modeled them using an one-zone leptonic model. The shape of the optical/UV component of the SEDs is dominated by accretion 
disk  emission in all the activity states. In the X-ray band, significant thermal emission from the hot corona is inferred during
quiescent and first flaring states, however, during subsequent flares, non-thermal jet component dominates. The $\gamma$-ray emission in all the states can be well explained by inverse-Compton scattering of accretion disk photons reprocessed by the broad line region. 
 The source showed violent intra-night optical variability, coinciding 
with one of the high $\gamma$-ray activity states. An analysis of the overall X-ray spectrum fitted with an absorbed power-law 
plus relativistic reflection component hints for the presence of Fe K-$\alpha$ line and returns a high black hole spin value of $a$=0.96~$\pm$~0.14. We argue that 1H 0323+342 possesses dual characteristics, akin to flat spectrum radio quasars (FSRQs) as well as radio-quiet NLSy1s, though at a low jet power regime compared to powerful FSRQs. 
\end{abstract}





\section{Introduction}\label{sec1}
The advent of {\it Fermi Gamma-ray Space Telescope} (hereafter, {\it Fermi}), has changed our understanding of the high energy $\gamma$-ray sky. 
Majority of the $\gamma$-ray sources detected by {\it Fermi} are blazars \citep{2012ApJS..199...31N}. Interestingly, besides blazars, {\it Fermi} has also detected variable $\gamma$-ray emission from five radio-loud narrow line Seyfert 1 (RL-NLSy1) galaxies with high significance \citep{2009ApJ...699..976A,2009ApJ...707L.142A,2012MNRAS.426..317D,2011MNRAS.413.2365C}. A few more NLSy1 galaxies are suspected to be $\gamma$-ray emitters, 
although their detection significance is still low \citep{2011nlsg.confE..24F}. NLSy1 galaxies are a class of active galactic nuclei (AGN) 
with peculiar properties: their optical spectra are similar to conventional broad line Seyfert 1 galaxies, however, they have narrow Balmer lines (FWHM (H$_{\beta}$) $<$ 2000 km s$^{-1}$), weak [O~{\sc iii}] ([O~{\sc iii}]/H$_{\beta} <$ 3) and strong optical Fe~{\sc ii} lines \citep{1985ApJ...297..166O, 1989ApJ...342..224G}. They also have steep soft X-ray spectra \citep{1996A&A...305...53B, 1996A&A...309...81W,1999ApJS..125..317L} and show rapid X-ray flux variations \citep{1995MNRAS.277L...5P,1999ApJS..125..297L}. These observational characteristics are attributed to them having low-mass black holes ($\sim 10^{6}-10^{8} M_{\odot}$) accreting close to the Eddington limit \citep{2000ApJ...542..161P, 2000NewAR..44..419H,2004ApJ...606L..41G,2006ApJS..166..128Z, 2012AJ....143...83X}. However, \citet{2013MNRAS.431..210C}, using multi-wavelength data, have shown that RL-NLSy1 galaxies have black hole masses similar to blazars. These RL-NLSy1 galaxies comprising about 7\% of NLSy1 galaxy population, exhibit compact core-jet structure, flat/inverted radio spectra, high brightness temperature and superluminal patterns \citep{2006AJ....132..531K,2006PASJ...58..829D}.  Recently, kiloparsec-scale radio structures have been found in six RL-NLSy1 galaxies (\citealt{2012ApJ...760...41D}). 
Moreover, it has been recently reported that the optical and infra-red (IR) flux variations of some of these $\gamma$-ray emitting
RL-NLSy1 ($\gamma$-NLSy1) galaxies are similar to blazars (\citealt{2010ApJ...715L.113L,2013MNRAS.428.2450P,2012ApJ...759L..31J}). 
They also show the double hump structure in their broad-band spectral energy distribution (SED) 
\citep{2013ApJ...768...52P,2009ApJ...707L.142A,2012MNRAS.426..317D}. These observed properties therefore clearly indicate that 
$\gamma$-NLSy1 galaxies can host relativistic jets similar to the blazar class of AGN. However, NLSy1 galaxies are believed to reside in 
spiral hosts compared to blazars that are hosted by ellipticals \citep{2009arXiv0909.2576M}.

1H 0323+342 ($z$ = 0.063) is one among the five $\gamma$-NLSy1 galaxies detected by {\it Fermi}. This source is radio-loud (R = 318, where R is the radio 
loudness parameter defined as the ratio of 5 GHz to optical B-band flux densities; \citealt{2011nlsg.confE..24F}) and has a flat radio spectrum ($\alpha_r$ = 0.35, calculated using the 6 and 20 cm flux densities given in \citealt{2010A&A...518A..10V}; $S_{\nu} \propto \nu^{-\alpha}$). The very long baseline interferometry imaging observations have revealed one sided jet on pc scales (\citealt{2005AJ....130.1389L} in the MOJAVE\footnote{http://www.physics.purdue.edu/astro/MOJAVE/} project), which can be interpreted  as a consequence of the Doppler-beaming effect if the jets are not intrinsically asymmetric. However, a two-sided kilo-parsec 
scale radio structure has also been found, \citep{2008A&A...490..583A,2012ApJ...760...41D}, although a brightness temperature of T$_{b}$ $\sim$ 5 $\times$ 10$^{11}$ K \citep{2007ApJ...658L..13Z} indicates a small jet inclination. The optical spectrum of this source is similar to that of a conventional NLSy1 galaxy, having narrow H$_{\beta}$ line and weak [O~{\sc iii}]/H$_{\beta}$ ratio (FWHM$_{H_{\beta}}$ = 1520 km s$^{-1}$, [O~{\sc iii}]/H$_{\beta}$= 0.12; \citealt{2007ApJ...658L..13Z}). Using the width and luminosity of the H$_{\beta}$ line and the empirical scaling relations  \citep{2005ApJ...630..122G}, \citet{2007ApJ...658L..13Z} has determined its black hole mass as 10$^{7} M_{\odot}$. Moreover, because of its unusual physical properties, they speculated that this source could host a NLSy1$-$blazar composite nucleus, even before the launch of {\it Fermi}. This is the only $\gamma$-NLSy1 galaxy which is included in the 70 months {\it Swift}-BAT catalog as well as significantly detected by {\it INTEGRAL} \citep{2013ApJS..207...19B,2011MNRAS.417.2426P}. Of particular interest, a TeV flare was claimed to be marginally detected at a significance level of $\sim$ 2.5-3.3 $\sigma$ on 2001 October 10 with a peak rate of 0.62~$\pm$~0.19 Crab \citep{2004ApJ...613..710F}. {\it Hubble Space Telescope} observations point the source
to have a spiral structure \citep{2007ApJ...658L..13Z}, though, recently \citet{2012AAS...22033507H} claimed its surface brightness distribution 
to be well fit with an elliptical profile. In addition, they have not seen any evidence of a separate bulge or disk and this supports the 
idea of a recent merger. Further, this source is found to be significantly variable within the night in optical band with amplitude of 
variability $>$ 3\% \citep{2013MNRAS.428.2450P}. In this work, we use multi-wavelength observations of 1H 0323+342 to characterize its 
physical properties and environment. 

The paper is organized as follows. In section~\ref{sec2}, we describe the data reduction procedure used, while section~\ref{sec3} is devoted to 
the results obtained. We discuss our findings in section~\ref{sec4} and summarize the results in section~\ref{sec5}. Throughout the work, we adopt $\Omega_{m}$=0.27, $\Omega_{\varLambda}$=0.73 and  Hubble constant {\it H}$_{0}$ = 71 km s$^{-1}$ Mpc$^{-1}$.

\section{Multi-wavelength Observations and Data Reduction}\label{sec2}
\subsection{{\it FERMI- Large Area Telescope}}\label{subsec1}
The {\it Fermi}-Large Area Telescope (LAT; \citealt{2009ApJ...697.1071A}) data used in this work was collected over the last five years of {\it Fermi} operation, from 2008 August 05 to 2013 September 15. Data analysis is done using {\tt ScienceTools v9r31p1} along with the use of post-launch instrument response functions (IRFs) P7SOURCE\_V6. In the energy range of 0.1--100 GeV, only SOURCE class events are selected. The maximum zenith angle is set to 100\textdegree{} so as to avoid contamination from Earth limb $\gamma$-rays. Spectral fitting is done using both unbinned as well as binned likelihood methods. A galactic diffuse emission component and an isotropic component are considered as background models, in order to extract the $\gamma$-ray signal. We use the same galactic component (gal\_2yearp7v6\_v0.fits) and the isotropic component, iso\_p7v6source.txt\footnote{http://fermi.gsfc.nasa.gov/ssc/data/access/lat/BackgroundModels.html} which is used in the second {\it Fermi}-LAT catalog (2FGL; \citealt{2012ApJS..199...31N}). The normalization of both components in the background model are left as a free parameter during the spectral fitting. Moreover, in order to search for the highest energy photon, we select only CLEAN class events along with the use of P7CLEAN\_V6 IRFs and use  iso\_p7v6clean.txt for background modeling. We use {\tt gtsrcprob} tool to determine the highest energy of the $\gamma$-ray photon.

The significance of the $\gamma$-ray signal is evaluated by means of the maximum-likelihood test statistic TS (= 2$\Delta$ log(likelihood)) between models with and without a point source at the position of the source of interest. We apply a binned likelihood method to the five year average analysis of LAT data and to generate $\gamma$-ray spectra over different time periods. To do this, we include all the point sources from the 
2FGL catalog that fall within 15\textdegree{} region of interest (RoI) of the source. We use both  power-law (PL$_{\gamma}$) and log parabola (LP) models 
to test the presence of a possible curvature in the $\gamma$-ray spectrum of the source in its different activity states. For sources lying 
within 7\textdegree{} of the source, all parameters except the scaling factor are left free, for sources lying between 7\textdegree{} to 14\textdegree{} only normalization factor is kept free, whereas for remaining sources lying outside 14\textdegree{}, all parameters are fixed to their 2FGL catalog values. Further, for variability and spectral analysis, where the time period under consideration is small, we remove the sources from the model having TS $<$ 25. A second maximum-likelihood analysis is then performed on the updated source model. This model is then used for 
generation of $\gamma$-ray light curve and spectrum. In case of non-convergence of likelihood fitting, we freeze all the parameters of the sources lying outside 7\textdegree{} from the center of RoI, to the values obtained from the average likelihood fitting. In case of further non-convergence, fitting is repeated again by fixing the photon indices of sources further 1\textdegree{} inside and this process is repeated till the analysis converge. Though the $\gamma$-ray spectrum of 1H 0323+342 shows a significant break or curvature (modeled by LP model in 2FGL catalog), weekly binned light curves are produced by using the PL$_{\gamma}$ model, as the statistical uncertainties on the PL$_{\gamma}$ indices are smaller than those obtained from more complex models. We consider the source to be detected if TS $>$ 9, which corresponds to $\sim$ 3$\sigma$ detection \citep{1996ApJ...461..396M}. For 1 $<$ TS $<$ 9, we calculate 2$\sigma$ upper limit by varying the flux of the source given by {\tt gtlike}, till TS reaches a value of 4 (see for e.g. \citealt{2010ApJS..188..405A}). For TS $<$ 1, we have not calculated upper limits. The systematic uncertainty in the flux measurement is energy dependent: it amounts to 10\% at 100 MeV, 5\% at 560 MeV and 10\% above 10 GeV\footnote{http://fermi.gsfc.nasa.gov/ssc/data/analysis/LAT\_caveats.html}. All the errors quoted here refer to 1$\sigma$ statistical uncertainties, unless otherwise specified.
\subsection{{\it SWIFT} (BAT, XRT, UVOT)}\label{subsec2}
The {\it Swift} satellite \citep{2004ApJ...611.1005G} has observed 1H 0323+342 more than 80 times with all three instruments: the 
Burst Alert Telescope (BAT; \citealt{2005SSRv..120..143B}, 15$-$150 keV), the X-ray Telescope (XRT; \citealt{2005SSRv..120..165B}, 0.2$-$10 keV) and the Ultraviolet/Optical Telescope (UVOT; \citealt{2005SSRv..120...95R}) which can observe in six filters, namely, V, B, U, UVW1, UVM2 and UVW2.

This source is significantly detected by the BAT instrument and included in the {\it Swift}-BAT 70 months hard X-ray catalog \citep{2013ApJS..207...19B}. It is also included in the {\it Swift}-BAT hard X-ray transient monitor list \citep{2013ApJS..209...14K}. We downloaded the publicly available 70 months averaged spectrum of 1H 0323+342 and model it with a power-law model. The extracted spectrum is used to build the SED but not for modeling the emission mechanism, as it is averaged over a long duration of time. Also, as a part of the ongoing \textquotedblleft Palermo BAT survey project\textquotedblright, {\it Swift}-BAT light curves and spectra in 15$-$150 keV band are available on request at {\it Istituto di Astrofisica Spaziale e Fisica Cosmica di Palermo} (IASF Palermo). We requested and obtained the two week binned light curve.  

We analyze the XRT data with standard procedures ({\tt xrtpipeline v.0.12.8}), filtering and screening criteria using the latest version of HEASOFT package (6.14) and calibration data base (CALDB) updated on 2013 September 11. We use the standard grade selections of 0-12 in the photon counting mode. Both light curves and spectrum files are generated using {\tt xrtgrblc} version 1.6 \footnote{http://heasarc.gsfc.nasa.gov/lheasoft/ftools/headas/xrtgrblc.html}. This task selects source and background circular regions according to the current count rate. In order to handle both piled-up observations and cases where the sources land on bad columns, vignetting and point spread function correction is handled using {\tt xrtlccorr}. \textsc{Xspec} \citep{1996ASPC..101...17A} is used to do the spectral fitting using the spectrum files generated by {\tt xrtgrblc} task. We bin the data using {\tt grppha} to have at least 20 counts per bin and use two models, namely absorbed power-law (PL$_{X}$) and absorbed broken power law (BPL) for fitting, while galactic column density is taken from \citet{2005A&A...440..775K}. The uncertainties are calculated at 90\% confidence level. Further, in order to study the Seyfert characteristics of the source, other than blazar properties, we separately combine all 
the XRT spectra covering the same period as that covered to generate 70 months {\it Swift}-BAT spectrum and fit with a power-law plus a relativistic reflection model.

The {\it Swift}-UVOT data are integrated with {\tt uvotimsum} and then analyzed with {\tt uvotsource} task. The source region is chosen as a circle of 5$\arcsec$ for optical filters and 10$\arcsec$ for UV filters, centered at the source location, while 1$\arcmin$ sized background region is extracted from nearby source-free region. The observed magnitudes are de-reddened using the galactic extinction of \citet{2011ApJ...737..103S} and converted to flux units using the zero points and conversion factors of the {\it Swift}-CALDB \citep{2008MNRAS.383..627P}.

\subsection{Catalina Survey Data}\label{subsec3}
The Catalina Real-time Transient Survey (CRTS\footnote{http://nesssi.cacr.caltech.edu/DataRelease/}) is an ongoing program devoted to the search for optical transients in the V-band. The details of the data analysis procedure are given in \citet{2009ApJ...696..870D}. We have used the publicly available data of 1H 0323+342  which we correct for galactic extinction and further convert to flux units using the zero point flux given in \citet{1998A&A...333..231B}.

\subsection{Ovens Valley Radio Observatory Data}\label{subsec4}
For comparison with the light curves in other wavelengths, we consider the 15 GHz radio data obtained at Ovens Valley Radio Observatory (OVRO) as a part of its ongoing blazar monitoring program. Details of the data analysis procedure can be seen in \citet{2011ApJS..194...29R}.

\subsection{Intra-night optical monitoring observations}\label{subsec5}

We observed 1H 0323+342 on three nights in the month of November and December 2012 as a part of our ongoing campaign on NLSy1 galaxies. Two nights of observations (2012 November 19 and 20, in R and B-bands respectively) were carried out on the recently commissioned 130-cm telescope \citep{2010ASInC...1..203S} located at Devasthal and operated by Aryabhatta Research Institute of Observational Sciences (ARIES), India. The details of the instrument can be found in \citet{2013MNRAS.428.2450P}. A third night of observation was done in the R-band, on 2012 December 9, using the 2m Himalayan Chandra Telescope (HCT) at Indian Astronomical Observatory, located at Hanle, India. This telescope is of the Ritchey-Chr$\grave{e}$tien design 
with a f/9 beam at the Cassegrain focus\footnote{http://www.iiap.res.in/centers/iao}. The detector is a cryogenically cooled 2048 $\times$ 4096 chip, of which the central 2048 $\times$ 2048 pixels are used for imaging. The pixel size is 15$\mu$m$^{2}$, so that the image scale of 0.29$\arcsec$/pixel covers an area of 10$\arcmin$x10$\arcmin$ on the sky. The readout noise of CCD is 4.87 e$^{-}$/pixel and the gain is 1.22 e$^{-}$/ADU. The exposure time was appropriately chosen in order to get good signal to noise ratio.  The pre-processing of the images (bias subtraction, flat-fielding and cosmic ray removal) is done by applying the standard procedures in {\tt IRAF}\footnote{http://iraf.noao.edu/}. The instrumental magnitudes of the target and comparison stars in the image frames are determined by aperture photometry, using {\tt APPHOT}. We use the same comparison stars which we used in our earlier work \citep{2013MNRAS.428.2450P}. We then generate differential light curves (DLCs) for source$-$star and star$-$star pairs.

\section{Results}\label{sec3}

\subsection{Average {\it Fermi}-LAT Analysis}\label{subsec6}
We fit the five year {\it Fermi}-LAT data of 1H 0323+342 with both PL$_{\gamma}$ and LP model. Following \citet{2012ApJS..199...31N}, we use a 
likelihood ratio test and calculate the curvature of test statistic, TS$_{curve}$ = 2(log L$_{LP}-$log L$_{PL_{\gamma}}$). We 
find TS$_{curve}$ = 42.16 ($\sim$ 6$\sigma$) and thus conclude that a significant curvature is present in the $\gamma$-ray spectrum of the source. The results of the average $\gamma$-ray analysis is presented in Table~\ref{tab1}.

\subsection{Multi-wavelength temporal properties}\label{subsec7}
\begin{figure*}
\begin{center}
\includegraphics[height=16cm,width=18cm]{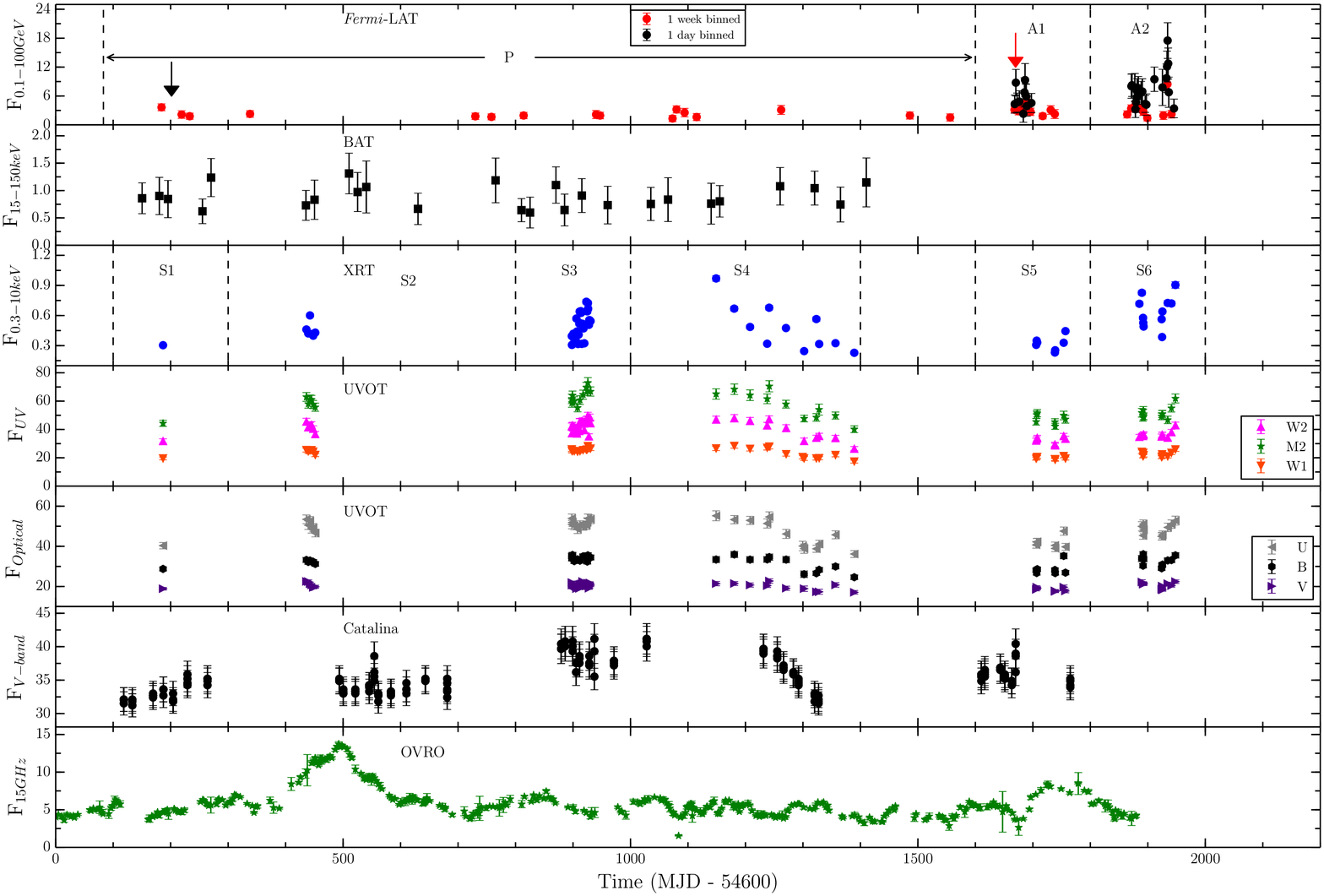}
\end{center}
\caption{Multi-band light curves of 1H 0323+342. The unit of OVRO data is 10$^{-14}$ erg cm$^{-2}$ s$^{-1}$ . Catalina and UVOT data are in 
units of 10$^{-12}$ erg cm$^{-2}$ s$^{-1}$. To show the variations, U, B, M2 and W2 band data points are shifted appropriately. XRT and BAT data points 
are in units of counts s$^{-1}$ and counts s$^{-1}$ pixel$^{-1}$ respectively, while {\it Fermi}-LAT $\gamma$-ray data are in units 
of 10$^{-7}$ photons cm$^{-2}$ s$^{-1}$. Black downward arrow shows the time of arrival of 32 GeV photon while gray downward arrow 
(red in online journal) shows the time of high INOV. P, A1 and A2 are the periods for which we generated $\gamma$-ray spectrum, to search 
for a possible curvature in the spectra. Anomalous variability behavior at different time periods is marked in the third panel from top. See text for details.}
\label{fig1}
\end{figure*}

The long-term multi-frequency light curves of 1H 0323+342, covering from 2008 August 05 to 2013 
September 15 is generated and is shown in Figure~\ref{fig1}. {\it Fermi}-LAT data is binned weekly, however, during the $\gamma$-ray
flaring period, one-day binned light curve is also generated. {\it Swift}-BAT data are binned over 15 days and we show only those fluxes 
which are at least 2$\sigma$ significant. {\it Swift}-XRT/UVOT observations correspond to one point per observation id (obsId). It is 
quite evident from the light curves that the source has shown anomalous variability behavior. However, this cannot be statistically claimed 
from any correlation studies due to the sparseness of the data.  From visual inspection, no correlation between radio and 
$\gamma$-ray light curves is observed and in addition, many X-ray and optical flares do not have $\gamma$-ray counterparts. To 
gain more insight into this anomalous variability behavior, we divide the {\it Swift} observations into six different time periods (S1 $-$ S6)  as shown in the third panel of Figure~\ref{fig1}. The source is found to be nearly stable during the periods S1 and S2. During S3 and S4, while significant variations are noted in optical and X-ray bands, no such variability is seen in the $\gamma$-ray band. However, during the periods S5 and S6, 
consistent flux variations are seen in all wave-bands. In S4 and S6 periods, the source
has similar X-ray peak flux, however, the corresponding $\gamma$-ray flux  during the later period is much higher when compared with the former.  We interpret this anomalous behavior of the optical/X-ray and $\gamma$-ray light curve as a result of variation in the jet environment associated with
the location of the emission region. {\it Swift}-XRT observed maximum flux from 1H 0323+342 on MJD 55749 (obsId 00036533019), when the detected flux is 3.12$^{+0.22}_{-0.20}$ $\times$ 10$^{-11}$ erg cm$^{-2}$ s$^{-1}$ (using PL$_{X}$ model) corresponding to an isotropic X-ray luminosity of $\sim$ 3 $\times$ 10$^{44}$ erg s$^{-1}$ which is comparable to average $\gamma$-ray luminosity of the source.

Using the recipe of \citet{2003MNRAS.345.1271V}, we calculate the fractional root mean square variability amplitude ({$F_{var}$}) in different energy bands (Table~\ref{tab2}). This parameter is found to be maximum for the $\gamma$-ray band while no statistically significant variation is seen in the 15-150 keV light curve. Flux variation is not seen in the hard X-ray band owing to the poor sensitivity of {\it Swift}-BAT in smaller time bins. The optical/UV light curves have minimum {$F_{var}$} whereas for soft X-ray light curve this parameter lies between optical and $\gamma$-rays. The radio emission can be due to a superposition of many jet components \citep{1981ApJ...243..700K} and thus the variations in the radio band might not be directly related to the variations in other wavebands. Barring the radio and hard X-ray bands, the amplitude of variability is found to increase with frequency, similar to blazars.

\begin{figure*}
\begin{center}
\includegraphics[height=8cm,width=12cm]{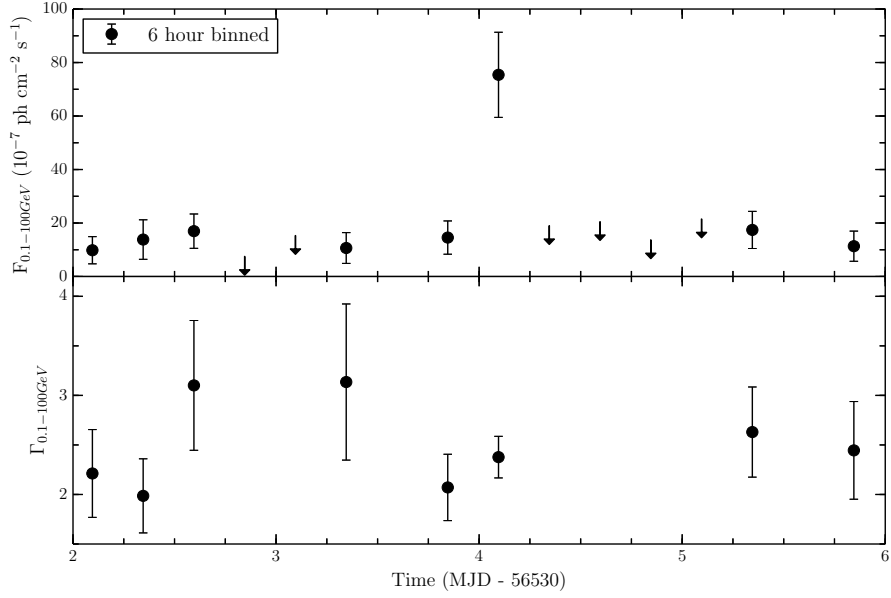}
\end{center}
\caption{Gamma-ray light curve of 1H 0323+342 during the period of high activity (MJD 56532-56536). Upper limits are shown by downward arrows.}
\label{fig2}
\end{figure*}

The recent GeV flare detected from 1H 0323+342 \citep{2013ATel.5344....1C} has its daily $\gamma$-ray flux as high as 
(1.80 $\pm$ 0.37) $\times$10$^{-6}$ photons cm$^{-2}$ s$^{-1}$. The photon statistics during this epoch is good enough to generate a six hour binned light curve describing this high activity period from 2013 August 28 to 2013 September 1 (see Figure~\ref{fig2}). Contrary to earlier observations, where the high amplitude $\gamma$-ray variability of $\gamma$-NLSy1 galaxies were characterized by longer timescales ($\geqslant$ 1 day; see for e.g., \citealt{2011MNRAS.413.2365C,2012A&A...548A.106F}), here the detected flux increases from $F_{1}$ = (1.46 $\pm$ 0.62) $\times$ 10$^{-6}$ photons cm$^{-2}$ s$^{-1}$ to $F_{2}$ = (7.54 $\pm$ 1.59) $\times$ 10$^{-6}$ photons cm$^{-2}$ s$^{-1}$ within $\Delta$t = 6 hours. We calculate a flux doubling timescale of $\tau_{d}$ = 2.53 $\pm$ 0.73 and $\tau_{d}$ = 4.41 $\pm$ 1.49 hrs while assuming exponential and linear flux increase respectively \citep{1999ApJ...527..719Z}. From the analysis of the 
photon indices at various epochs, it emerges that photon index is relatively hard (2.38 $\pm$ 0.21) when the source is 
brightest, as compared to 2.79 $\pm$ 0.05 obtained from five years average analysis. The better photon statistics of this sudden rising phase permit us to go for further smaller time bins ($\sim$ 100 min). Though in most of the time bins there is hardly any significant detection, during a time bin (centered at MJD 56534.13) the peak $\gamma$-ray flux is measured as (9.15 $\pm$ 2.31) $\times$10$^{-6}$ photons cm$^{-2}$ s$^{-1}$ with photon index of 2.34 $\pm$ 0.25 (TS = 76) and 24 counts are registered. This flux, in turn equals to the isotropic $\gamma$-ray luminosity (L$_{\gamma}$) = 4.7 $\times$ 10$^{46}$ erg s$^{-1}$, which is almost 160 times greater than its average $\gamma$-ray luminosity. Following 
\citet{2011A&A...530A..77F}, we select time bins equal to good time interval (GTI) to search for smaller scale variability. 
Of the twelve GTI bins (between MJD 56533.75 to 56534.25), there are only two significant detection and again, as in 6 hour binned light curve, the maximum occurred on MJD 56534.13 (GTI size $\sim$ 25 minute). This indicates the detection of a rapid $\gamma$-ray flare lasting for 
less than 25 minutes. However, as shown in Figure~\ref{fig2}, this flare is unresolved at 6 hours binning and even at GTI scale. To our knowledge such rapid $\gamma$-ray flare is observed for the first time from a $\gamma$-NLSy1 galaxy.

\subsection{Gamma-ray spectral analysis}\label{subsec8}
In the 2FGL catalog, 1H 0323+342 is modeled using a LP which indicated for the presence of  curvature in its $\gamma$-ray spectrum. To study 
the $\gamma$-ray spectrum during different activity states of the source, we select three different time periods, a quiescent and two 
relatively active states (P, A1 and A2 respectively; see Figure~\ref{fig1}). We apply both PL$_{\gamma}$ and LP models, use binned likelihood method for fitting and the results are given in Table~\ref{tab3}. The quiescent state (P) clearly indicates for the presence of a curvature (TS$_{curve}$ = 30.5 $\approx$ 5$\sigma$), however, no such curvature is found during the active states A1 and A2. This is in contrast to that of 
the $\gamma$-NLSy1 galaxy SBS 0846+513, where a significant spectral curvature is reported during a flaring state \citep{2013MNRAS.436..191D}.

\begin{figure*}
\begin{center}
\includegraphics[height=7cm,width=18cm]{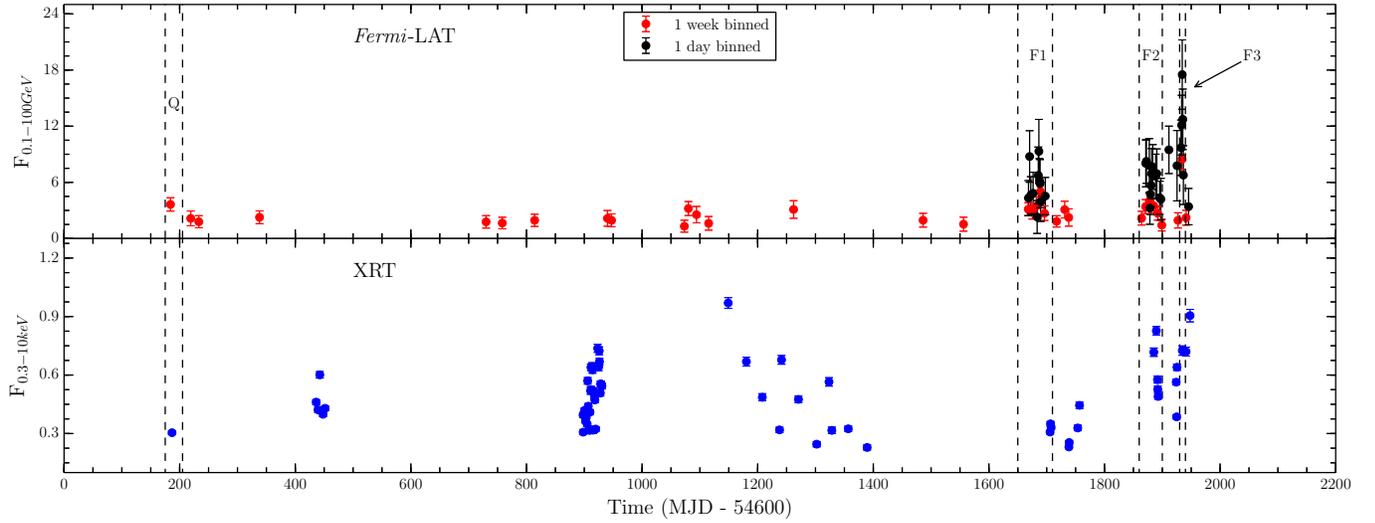}
\end{center}
\caption{Different activity periods selected for SED generation and modeling. Symbol Q corresponds to a quiescent state while F1, F2 and F3 are three flaring states when enhanced $\gamma$-ray emission is detected from the source. All units are same as in Figure \ref{fig1}.}
\label{fig3}
\end{figure*}

\begin{figure*}
\hbox{
\includegraphics[width=6.0cm]{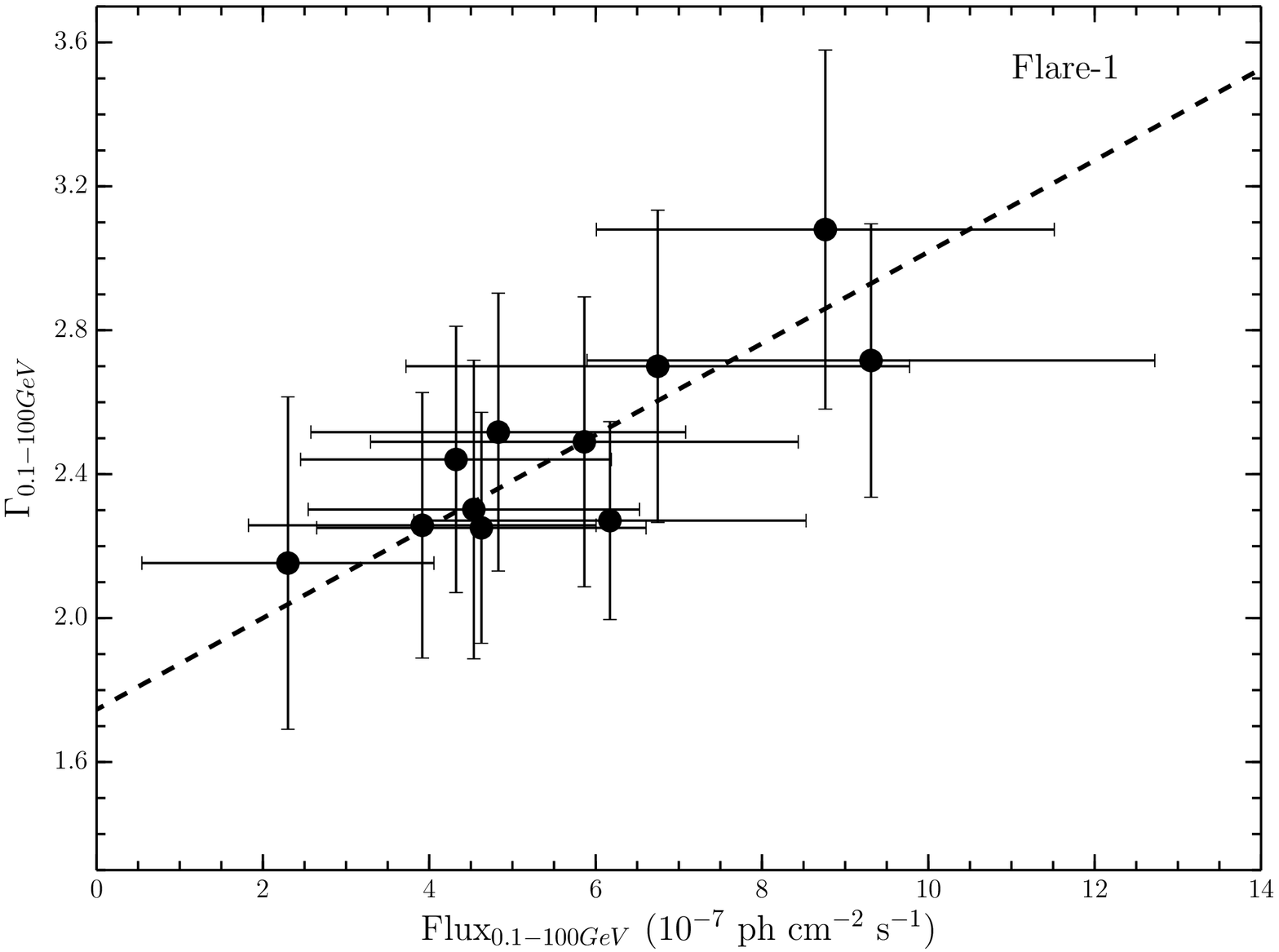}
\includegraphics[width=6.0cm]{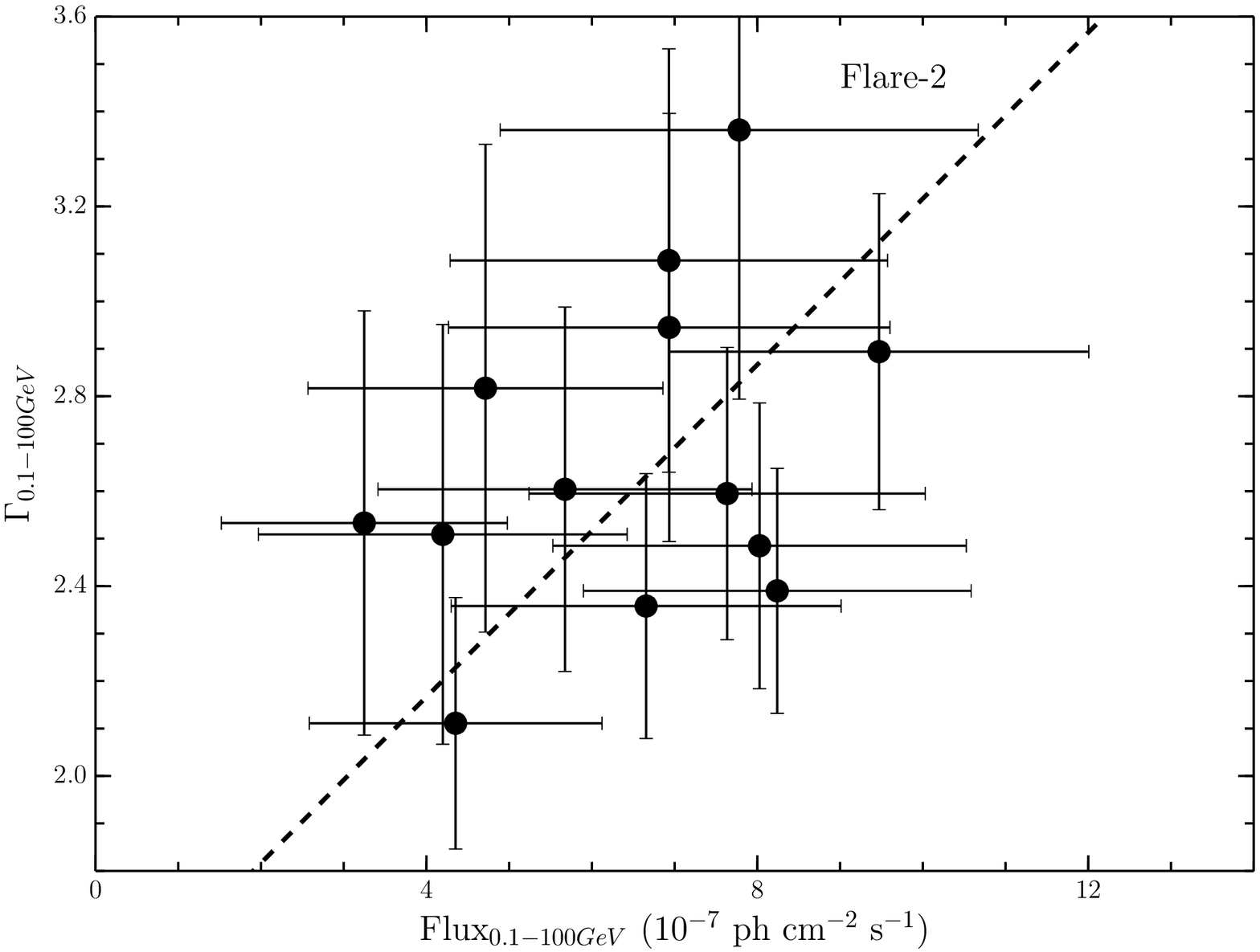}
\includegraphics[width=6.0cm]{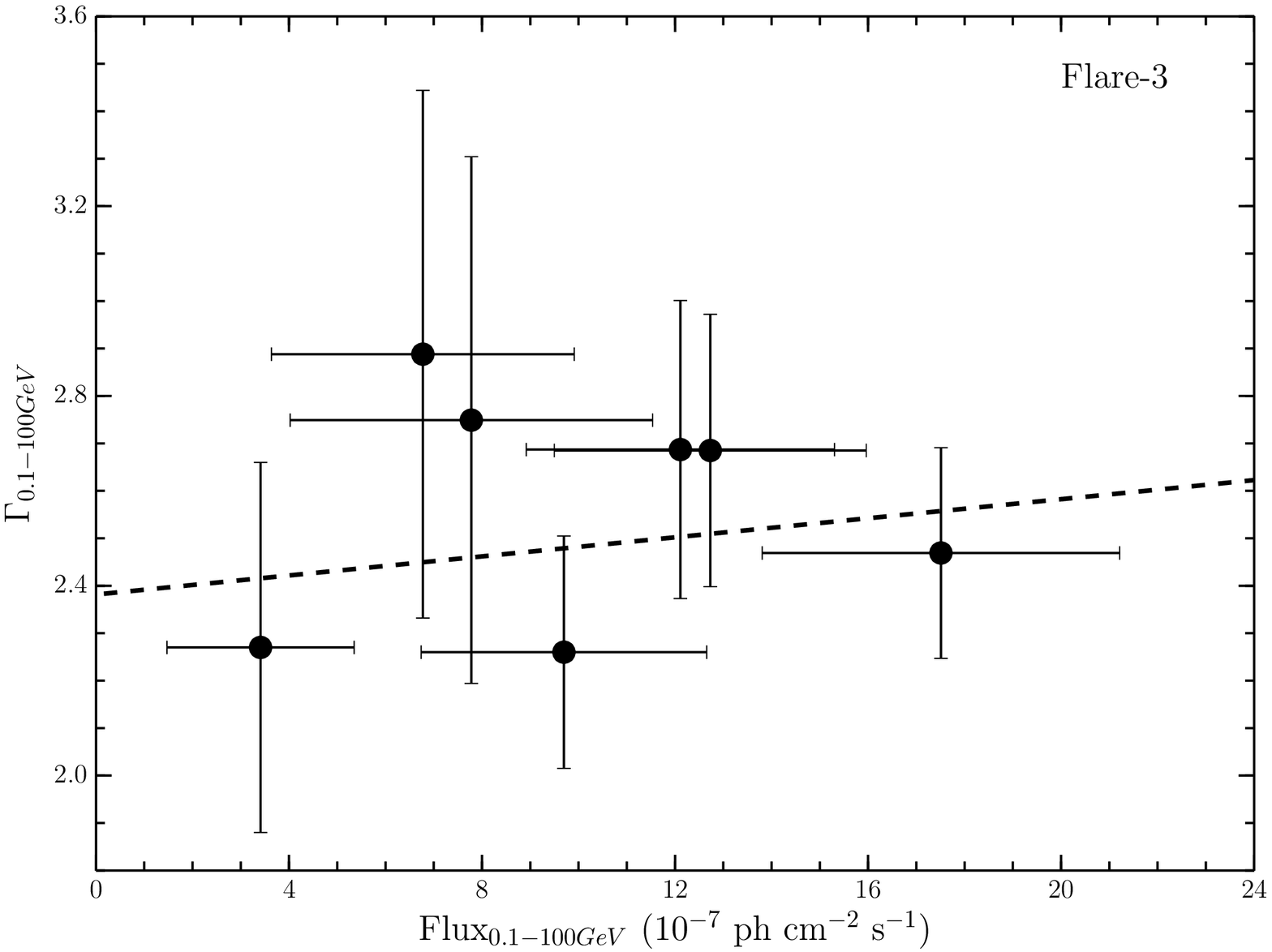}
}
\caption{Photon flux vs. photon index variations of 1H 0323+342 during different flaring states. Dashed line is the weighted linear least squares 
fit to the data.}
\label{fig4}
\end{figure*}

For SED analysis, we select four different time periods corresponding to different $\gamma$-ray activity states of the source 
(Q, F1, F2 and F3; Figure~\ref{fig3}). We find the PL$_{\gamma}$ to be a good representation of the $\gamma$-ray spectrum in all four 
different activity states. We note that the presence of curvature in the P-state but not in Q-state could be due to low photon statistics
of the Q-state.  Good photon statistics during these flaring episodes allow us to generate one day binned $\gamma$-ray light curve 
and thereby determine the variation of photon index with respect to the brightness of the source. The results are shown in 
Figure~\ref{fig4} along with a weighted linear least-squares fit to the data using {\tt fitexy} \citep{1992nrca.book.....P}. During F1 and F2 states, 
a clear \textquotedblleft softening when brightening\textquotedblright is seen, however, no significant correlation between photon index 
and flux is found during F3 state. 

\subsection{Energy of the highest energy gamma-ray photon}\label{subsec9}
We separately analyze the LAT data using event class CLEAN, to determine the energy of the highest energy photon detected from the source. In 
the first five years of {\it Fermi} operation, highest energy photon was detected on 2008 December 01 (black downward arrow in Figure~\ref{fig1}) at a distance of 0.04\textdegree{} from the source with an energy of 32.73 GeV (99.99\%probability of detection). This energy closely satisfies the transparency criteria of $\gamma-\gamma$ pair production with BLR photons \citep{2009MNRAS.397..985G}, and hence the emission region may lie inside the BLR. Further, we also search for highest energy photon during the different states considered for SED modeling (see Figure~\ref{fig3}). During 
the F1 state, the highest energy photon is having an energy of 3.13 GeV (98.61\% detection probability) arriving at 0.3\textdegree{} away from the 
location of 1H 0323+342, while during the F2 state, it is 4.68 GeV (99.60\% detection probability), detected at 0.15\textdegree{} away from the source. During F3 state, we find the energy of the highest energy photon to be 3.01 GeV (98.69\% probability) and detected at 0.29\textdegree{} 
away from the source.

\subsection{X-ray spectral analysis}\label{subsec10}
It is known that radio-loud AGN have flatter X-ray spectra compared to radio-quiet sources. Such hard X-ray spectra is an indication of
the presence of a relativistic non-thermal jet \citep{2000ApJ...531...52G,1999ApJS..125..317L,2014MNRAS.440L.111G}. To study the X-ray spectrum of 1H 0323+342, we use PL$_{X}$ and BPL models. Though, in most of the observations, PL$_{X}$ gives better fit, BPL is found to 
be a better representation of the data during a flaring state (F2) in the $\gamma$-ray band. During the quiescent state (Q in 
Figure~\ref{fig3}), X-ray spectrum is very well described by a simple PL$_{X}$ model with a soft PL$_{X}$ index of 1.95~$\pm$~0.06 which is similar to the average
value, $\Gamma_X$ = 2.00 $\pm$ 0.25 obtained for radio-quiet NLSy1 galaxies \citep{2000ApJ...531...52G}.  During the activity state F2, a PL$_{X}$ fit gives $\chi^{2}_{r}$ = 1.19 (231 {\it dof}) whereas BPL 
fit gives $\chi^{2}_{r}$ = 1.06 (229 {\it dof}). The probability that the fit improvements were by chance (obtained using the {\it F-test}) 
equals to 1.23 $\times$ 10$^{-6}$, hence strongly indicating the presence of a break in the X-ray spectrum. This result can be interpreted as, 
probably during quiescent $\gamma$-ray state, the X-ray emission is significantly 
contaminated by thermal corona emission \citep[resulting in a soft X-ray spectrum;][]{1999ApJS..125..317L}, while during the $\gamma$-ray outbursts, the power-law component 
from jet dominates \citep{2009AdSpR..43..889F}. In the third $\gamma$-ray flaring state (F3), X-ray spectrum is found to be better fitted by 
PL$_{X}$ model with a flat photon index value of 1.66 $\pm$ 0.06, \citep[similar to the photon indices of X-ray selected FSRQs;][]{2008MNRAS.391.1981M}. 
Such a flat photon index is a clear signature of the presence of non-thermal rising jetted emission which overwhelms the 
thermal coronal emission. Thus, a significant X-ray spectral evolution is found between the  quiescent and $\gamma$-ray flaring states of 1H 0323+342.

\begin{figure*}
\begin{center}
\includegraphics[width=10.0cm]{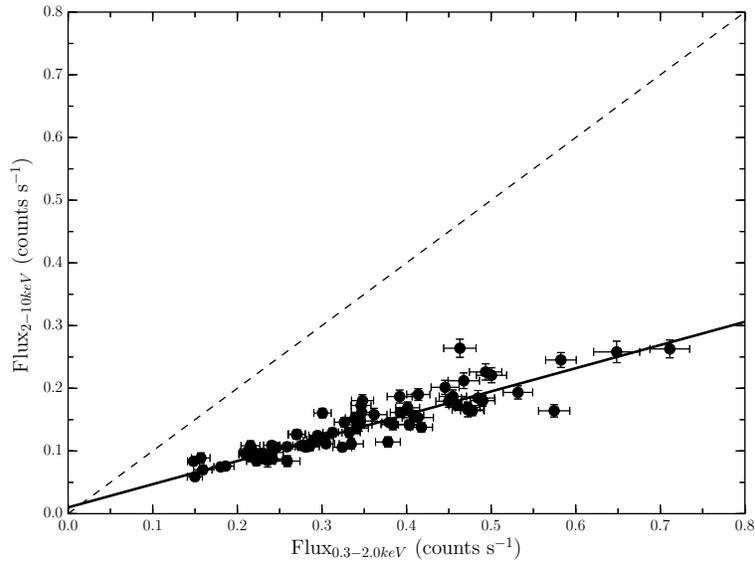}
\caption{Hard versus soft X-ray count rate plot of 1H 0323+342. The best-fitting linear model is represented by solid line, whereas the dashed line refers to the perfect one-to-one correlation. A \textquoteleft softer when brighter' trend is evident.}
\end{center}
\label{fig5}
\end{figure*}

\begin{figure*}
\centering
\includegraphics[width=10.0cm]{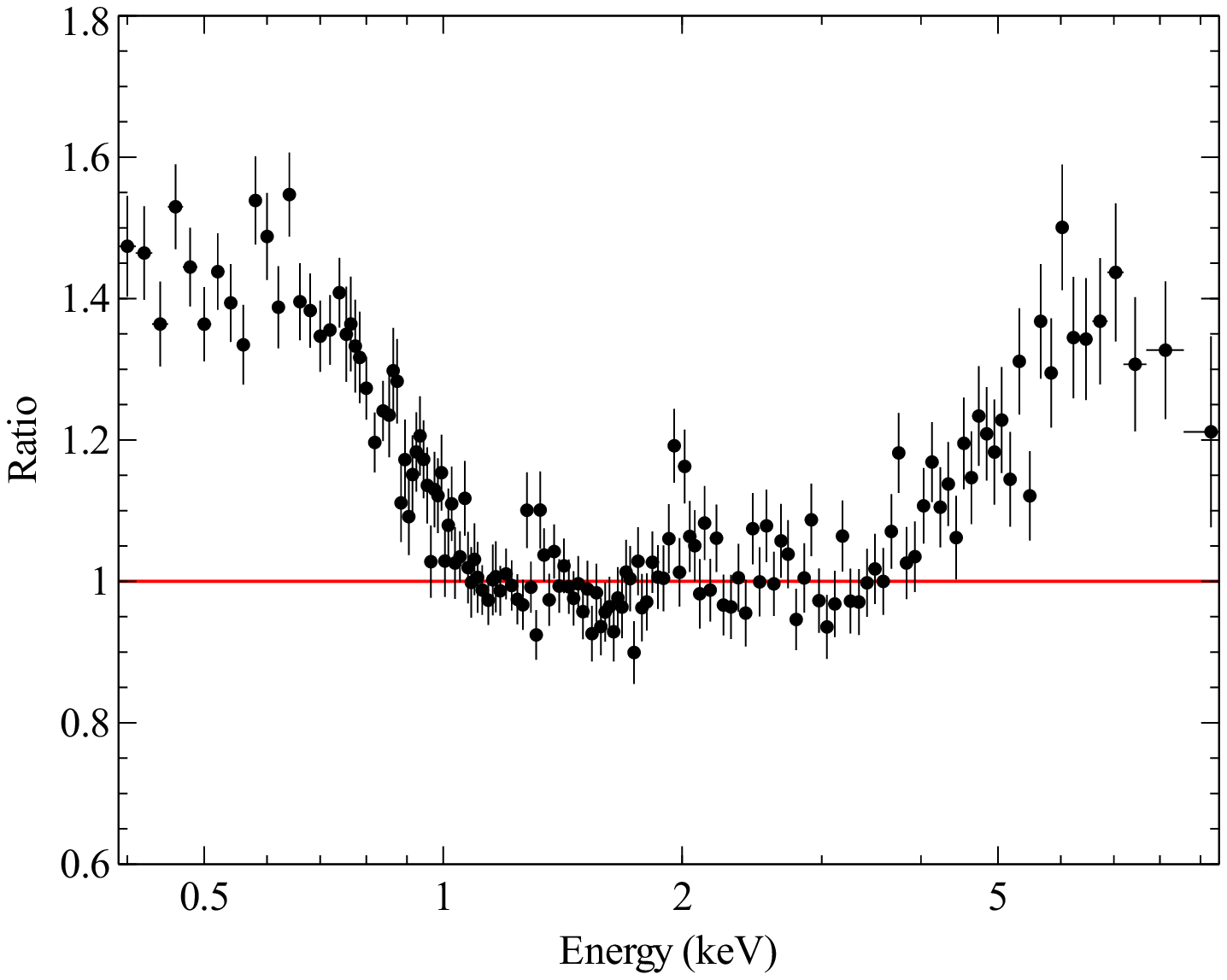}
\caption{Ratio of the average XRT spectrum to an absorbed power-law, fit between 1--4~keV. A soft excess is clearly visible at low energies, and a possible Fe line feature at high energies.}
\label{plratio}
\end{figure*}

To test the overall X-ray spectral variability of 1H 0323+342, we show in Figure~\ref{fig5} the plot of soft count rate (0.3--2 keV) versus hard count 
rate (2--10 keV, see for e.g., \citealt{2001MNRAS.321..759C}) using all XRT observations and perform a linear fit using the routine {\tt fitexy}. 
The solid line represents the best fit line corresponding to y = 0.37x + 0.01 whereas the dashed line indicates the one to one 
correlation (y=x). A \textquoteleft softer when brighter' spectral variability trend is evident. 
This X-ray spectral variation can be explained by the presence of 
a significant soft excess (see Figure~\ref{plratio}) whose variability behavior is not identical to the hard X-ray component. 
Similar  behavior is also recently noted in another RL-NLSy1 
galaxy PKS 0558$-$504 \citep{2013MNRAS.433.1709G}.

Figure~\ref{plratio} shows the ratio of the averaged XRT spectrum (total exposure $\sim$ 73 ksec after combining all XRT observations lying in the the same period which is used to generate {\it Swift}-BAT 70 months spectrum) to an absorbed power-law, fit between 1--4 keV. A soft excess is visible below $\sim$1~keV, and a possible iron line feature above $\sim$4~keV. The absorption is modeled using \textsc{TBabs} in \textsc{Xspec}, with a column density of $1.27\times10^{21}$~cm$^{-2}$, and the power-law index is found to be $2.02\pm0.06$.

\begin{figure*}
\centering
\includegraphics[width=10.0cm]{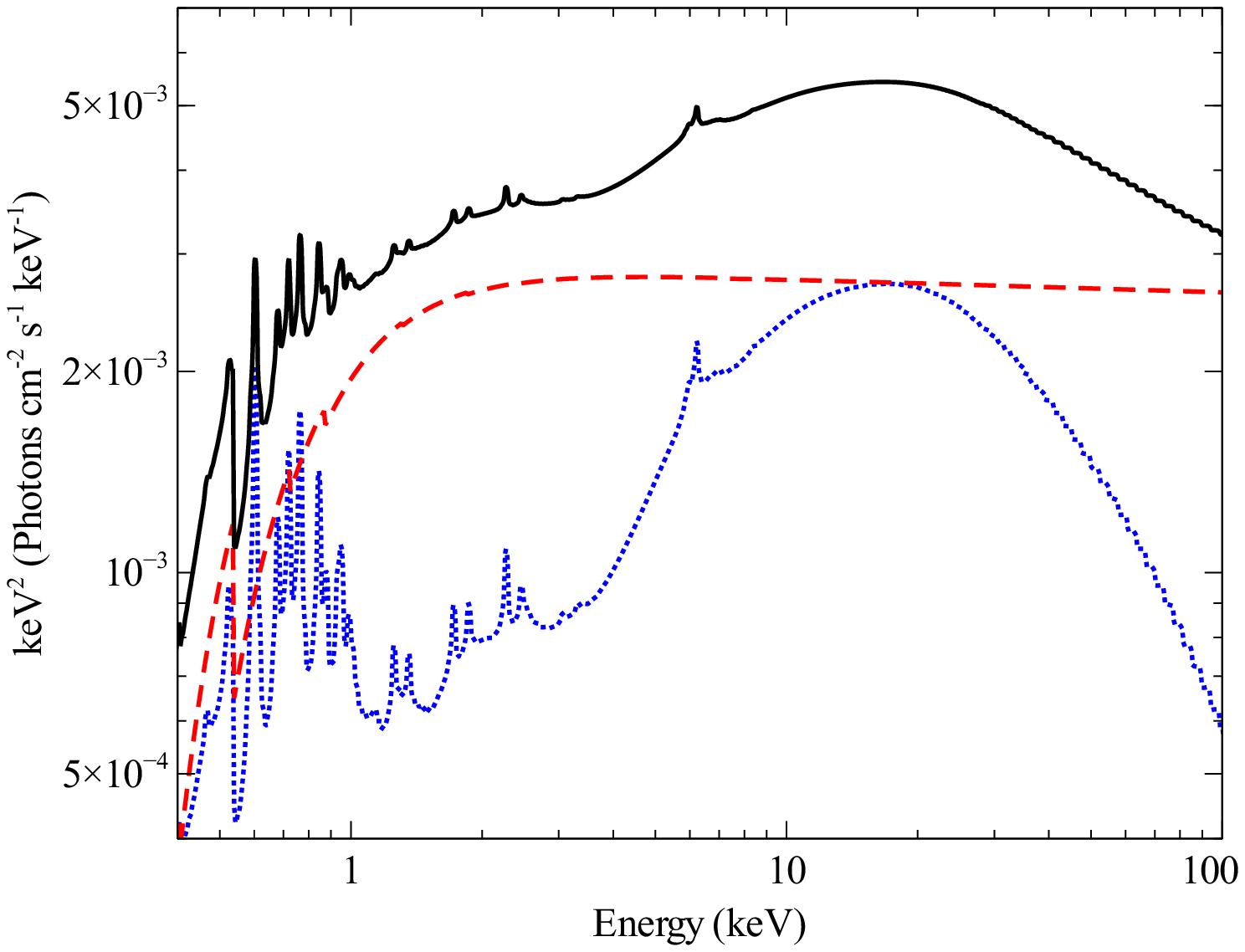}
\caption{Best fitting power-law plus relativistic reflection model. The black line shows the complete model, the red dashed line the power-law, and the blue dotted line the blurred reflection component. The blurring is less extreme in this object due to the low inclination of the disk, meaning that Doppler and boosting effects are limited.}
\label{refmodel}
\end{figure*}

We initially attempt to fit these features with an ionized reflection model, using \textsc{Reflionx} \citep{Ross05}. We find a best fit of $\chi^2_\nu=1.07$, for 587 degrees of freedom. We find an iron abundance of 2.9$\pm0.6$ relative to solar, a photon index of $1.94\pm0.06$, and an ionization parameter of $231\pm20$ erg cm s$^{-1}$. Convolving the reflection model with a relativistic blurring model \citep[\textsc{Relconv}, ][]{Dauser10} improves the fit to $\chi^2_\nu=1.04$, for 584 degrees of freedom, an improvement of $\Delta\chi^2=25$, for three additional degrees of freedom. We freeze the inclination at ten degrees and the break in the emissivity profile at 6 $R_\textrm{G}$, and find best fit parameters of $a=0.96\pm0.14$ for the spin, $\Gamma_X=2.02\pm0.06$ for the photon index, $\xi=212\pm12$ erg cm s$^{-1}$ for the ionization parameter, and $q\geq 7.5$ and $1.8\pm0.4$ for the inner and outer emissivity indices, respectively. The best fit model is shown in Figure~\ref{refmodel}. Fixing the inclination is likely to be responsible for the small error on the spin value obtained as these parameters are partially degenerate. However, the low inclination angle used is physically motivated, and the data quality is not sufficient to constrain both parameters. Fixing the radius of the break in the emissivity profile to 6~$R_\textrm{G}$ is reasonable, as this is around the value predicted for the break by simple models of the coronal geometry \citep[e.g.,][]{Wilkins12,Dauser13}. While still somewhat arbitrary, this choice should not affect parameters other than the emissivity indices, which will change to try and emulate the true profile.

\begin{figure*}
\centering
\includegraphics[width=10.0cm]{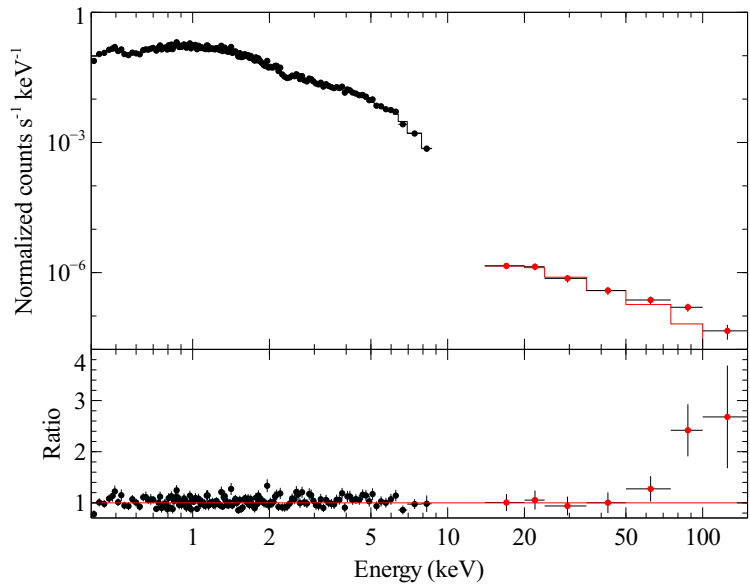}
\caption{Top: \emph{Swift} XRT (black) and BAT (red) data, fit with a power-law plus blurred reflection model from 0.4--50 keV, then extrapolated up to 150 keV. Bottom: Ratio of the data to this model. A significant excess is seen above 50 keV, which is likely to be due to the presence of the jet in the source.}
\label{fitdata}
\end{figure*}

We then extended this model to include the 70 months averaged BAT data, allowing for a difference in normalization between the two detectors, and restricting the XRT data to the same period as covered by the BAT. Fitting the model up to 50 keV gives a good fit, with a difference in normalization of $1.06\pm0.11$, but extrapolating to higher energies it under predicts the flux by a factor of $\sim3$, as shown in Figure~\ref{fitdata}. We assume that this upturn is due to emission from the jet. 

We also investigate individual spectra from different $\gamma$-ray flares, compared with the spectrum from a quiescent period (Q). The data quality is too low for detailed analysis, so we use a simple absorbed power-law model to investigate these spectra. Both the quiescent state (Q) and first flare (F1) show evidence for a broad excess around 6~keV.

Using the best fit model, and compensating for absorption, we find X-ray luminosities in the 0.3--10~keV, 2--10~keV, and 10--100~keV bands of $9.2\times10^{43}$~erg~s$^{-1}$, $2.1\times10^{44}$~erg~s$^{-1}$ and $1.9\times10^{44}$~erg~s$^{-1}$, respectively.

\subsection{Spectral energy distribution}\label{subsec11}
\begin{figure*}
\hbox{
\includegraphics[width=8.5cm]{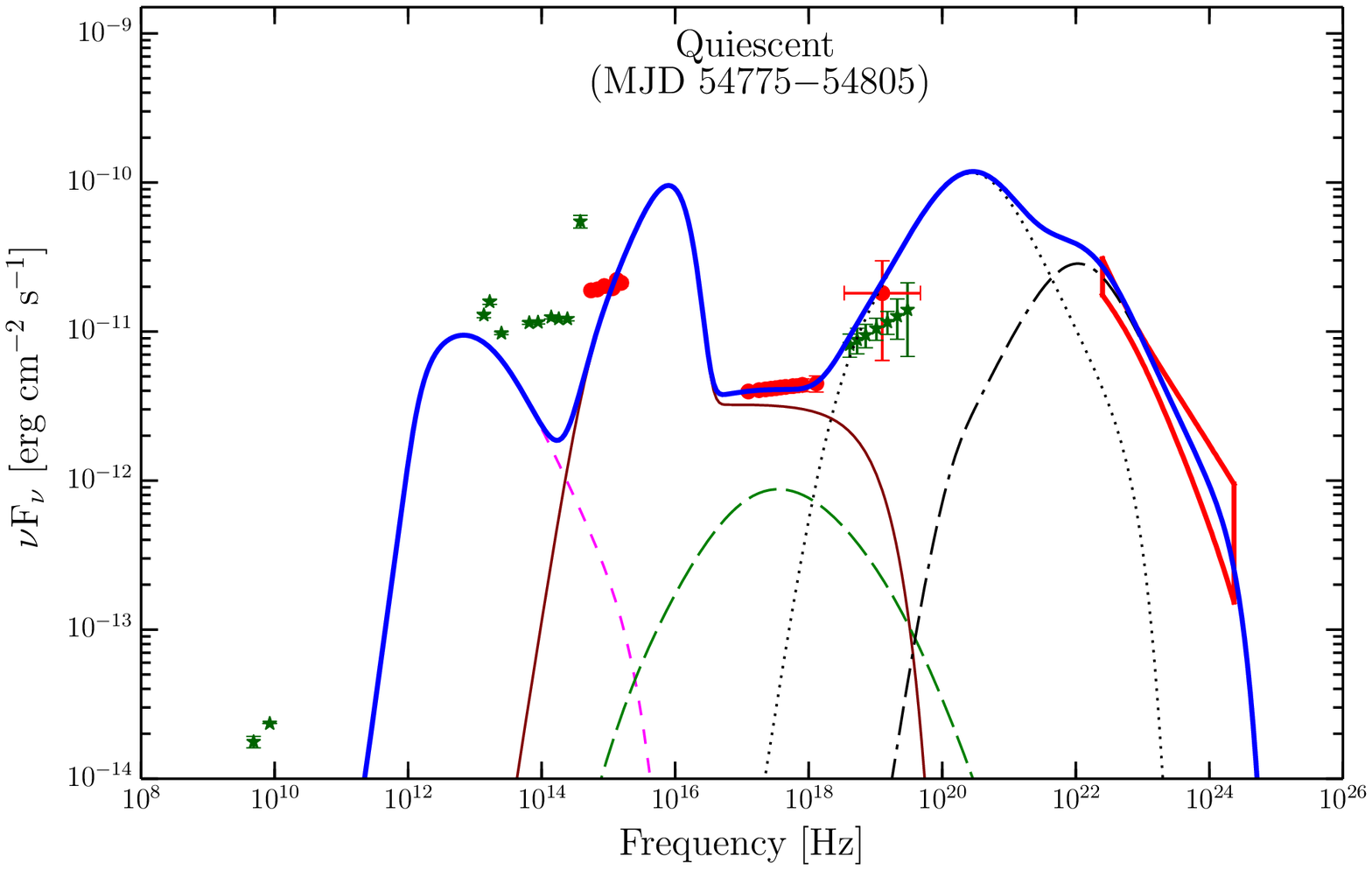}
\includegraphics[width=8.5cm]{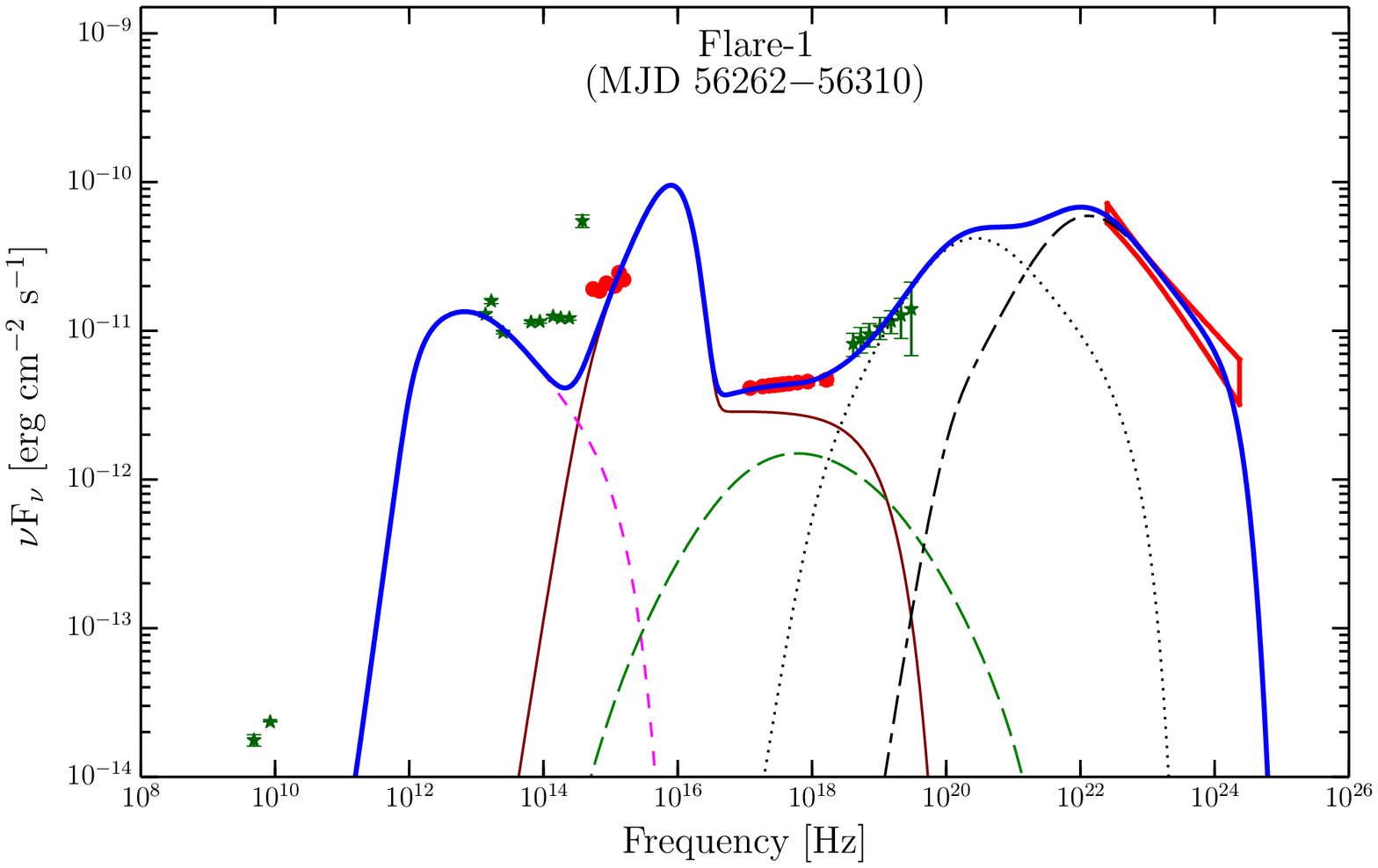}
}
\hbox{
\includegraphics[width=8.5cm]{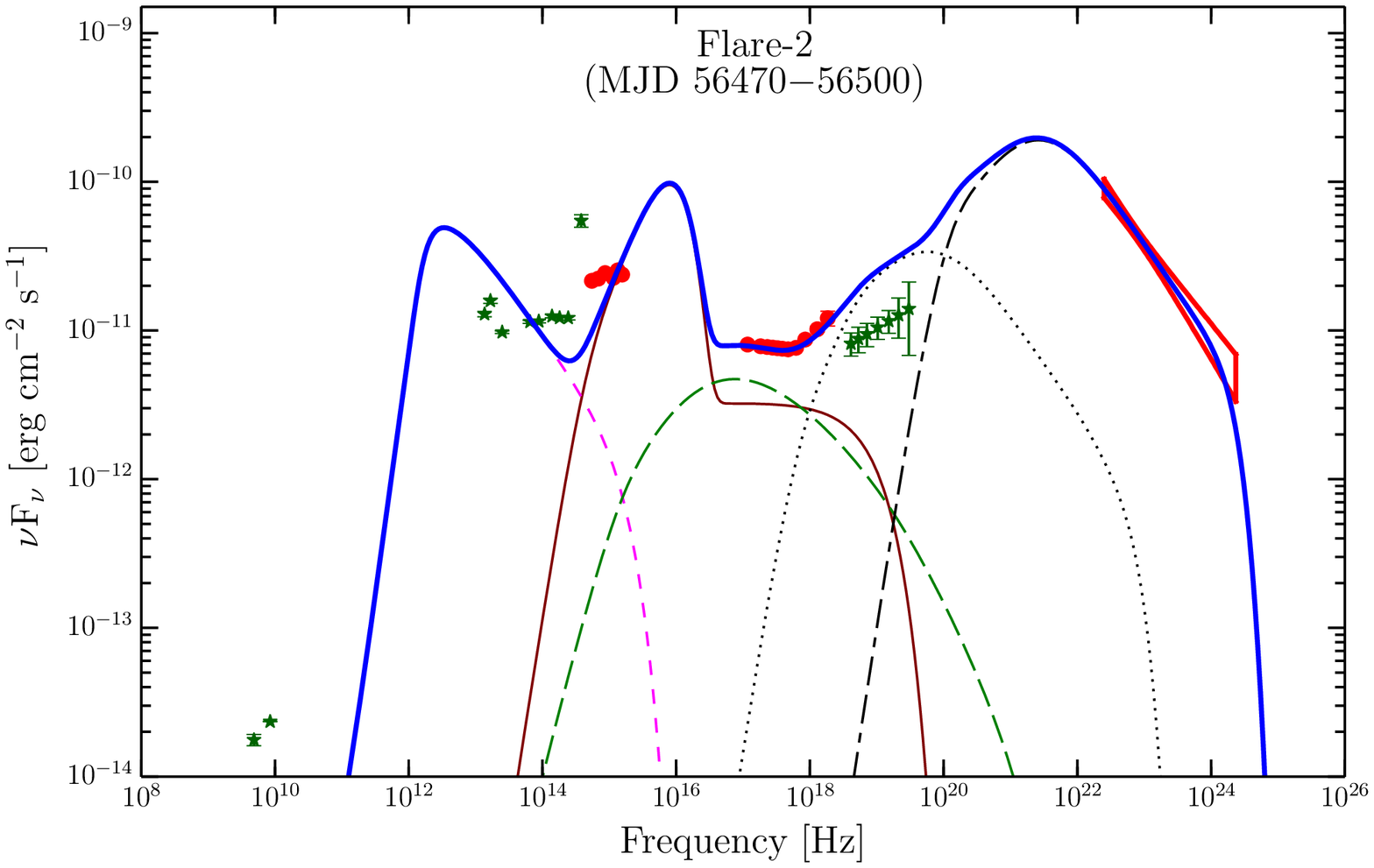}
\includegraphics[width=8.5cm]{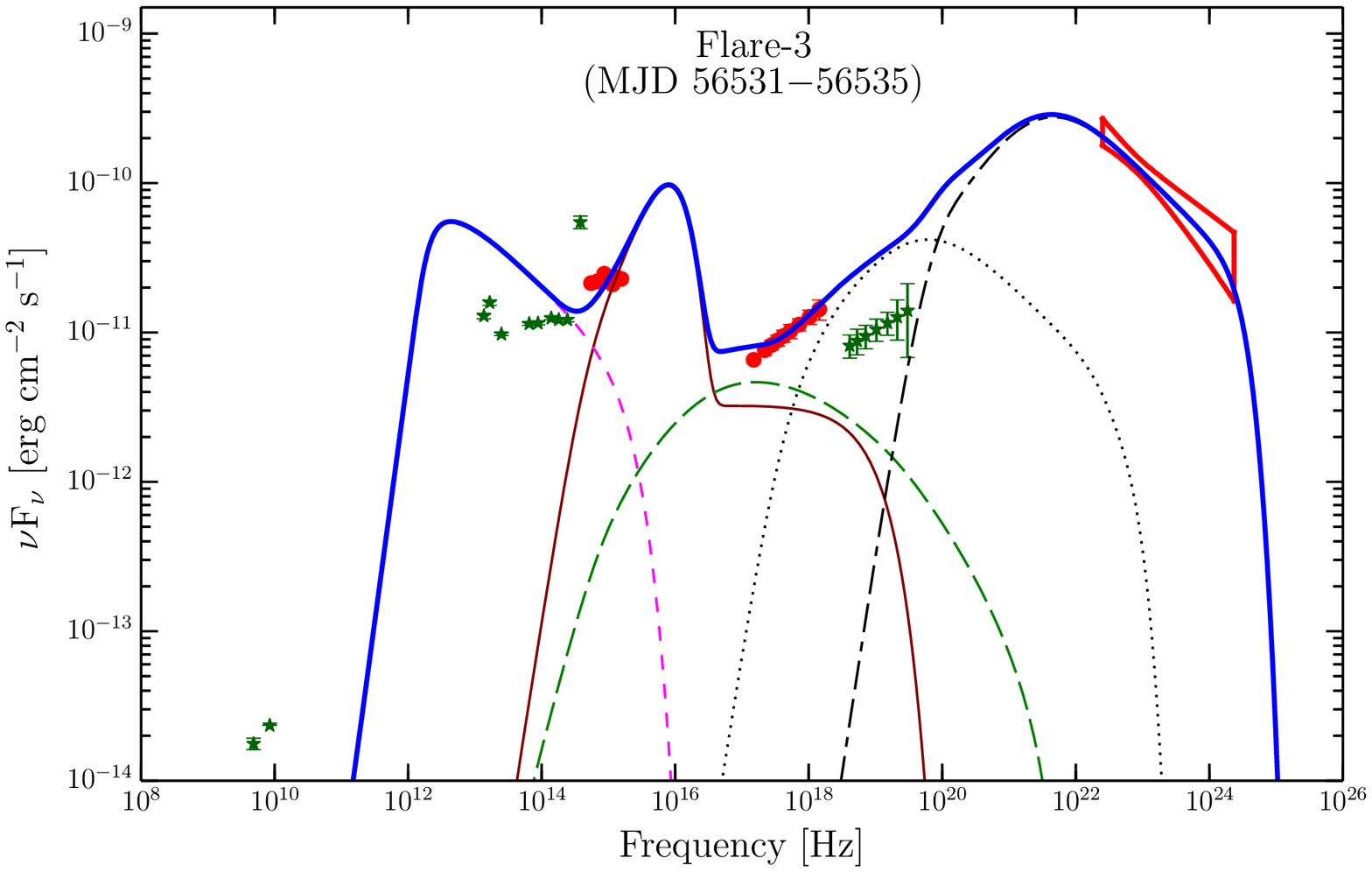}
}
\caption{SEDs of 1H 0323+342 during different activity states. Simultaneous {\it Swift}-UVOT, XRT (filled circles) and {\it Fermi}-LAT (bow-tie plot) observations are shown by red color. Archival data is shown with dark green color. In the quiescent state SED, we have simultaneous 15$-$150 keV {\it Swift}-BAT observation. We also show 70 months averaged {\it Swift}-BAT spectrum, though it is not used for modeling. Thin solid line (maroon color) shows thermal emission. The synchrotron (pink), SSC (green) EC-disk and EC-BLR (black) processes are shown with dashed, long dashed, dotted and dotted dashed lines, respectively. Blue continuous line is the sum of all radiative components.}
\label{fig6}
\end{figure*}

For generating the SEDs, we average the flux over each of the four time intervals considered for modeling
marked as Q, F1, F2 and F3 in Figure~\ref{fig3}. We use simultaneous UVOT observations corresponding to the XRT observations selected. The quiescent state is chosen when the source is in the faint state in all wavebands. The derived flux values for these four time periods are given in Table~\ref{tab4}. All the four SEDs are modeled using a modified single-zone leptonic emission model of \citet{2012MNRAS.419.1660S}. In this model, the emission region is assumed to be a spherical blob moving relativistically with bulk Lorentz factor $\Gamma$ at a small angle $\theta$ to the line of sight.  
Assuming a conical jet with a semi-opening
angle ($\phi$) of 0.1, we consider the emission region to have a size of radius (R) = $\phi R_{\rm diss}$, where $R_{\rm diss}$ is the 
distance of the emission region from the central black hole. 
The region is filled with relativistic electrons which is assumed to be a broken power law distribution with indices {\it p} and {\it q} before and after break energy ($\gamma_{b}$mc$^{2}$), respectively. The electrons lose their energy through synchrotron emission in a randomly oriented magnetic field and inverse Compton scattering of synchrotron photons (SSC) and the photons external to the jet (EC). The magnetic field is considered to be in equipartition with relativistic particle distribution. The model is modified to include the emission from an accretion disk, X-ray corona, and EC scattering of disk photons (EC-disk) as well as disk photons reprocessed by 
the BLR (EC-BLR). A multi-temperature black body spectrum is assumed for the accretion disk emission \citep{2002apa..book.....F} whose inner and outer radii
are considered as 3$R_{\rm S}$ and  500 $R_{\rm S}$, where $R_{\rm S}$ is the Schwarzschild radius. For the corona we assume a flat power-law with an energy index equal to unity extending from 0.05 keV to 50 keV. The BLR is assumed to be a black body type spectrum peaking at $\sim$ 3.7 $\times$ 10$^{15}$ Hz (corresponding to rest-frame Lyman-alpha line frequency) from a spherical shell \citep{2008MNRAS.386..945T} with the size constrained by the empirical relations \citep{2007ApJ...659..997K,2009ApJ...697..160B}.  
The BLR is assumed to re-process 10\% of the disk luminosity \citep{2009MNRAS.397..985G}. 
The kinetic power of the jet is calculated by assuming both protons and electrons to have equal number densities. The protons are
assumed to be cold and thus contributes only to the inertia of the jet. The main parameters governing the SEDs can be obtained using the observed information available in optical/X-ray and $\gamma$-ray energies. The results of the SED modeling are shown in Figure~\ref{fig6} and the parameters are given in Table~\ref{tab5}. We have assumed the black hole mass as 2 $\times$ 10$^{7}$ $M_{\odot}$. In all states, the optical/UV component of the SEDs is dominated by thermal emission from the accretion disk whereas the $\gamma$-ray emission can be well explained by EC scattering of disk photons reprocessed by the BLR. This suggests that the location of the emission region is within the BLR.

In the quiescent (Q) and first flaring (F1) states, a significant contribution from the X-ray corona is observed. Further, during the last two flaring states (F2 and F3), the X-ray spectrum is dominated by non-thermal jet component over corona emission. The X-ray spectrum of the second flare (F2) exhibits a break which can be explained as a dominance of non-thermal jetted emission (primarily EC-disk) in hard X-ray band, whereas soft X-ray emission can be attributed to a combination of SSC and corona radiation. In the third flare (F3), the X-ray spectrum becomes very hard, predominantly by SSC plus EC-disk process. We also note a hardening of $\gamma$-ray spectrum in brighter state, when comparing different activity periods.

Some of the parameters obtained from our modeling differ from that obtained by \citet{2009ApJ...707L.142A} and this could be due to the fact that we have used simultaneous data whereas the data used in \citet{2009ApJ...707L.142A} is non-contemporaneous.

\subsection{Intra-night optical variability (INOV)}\label{subsec12}
\begin{figure*}
\hbox{
\includegraphics[width=6.0cm,height=12.0cm]{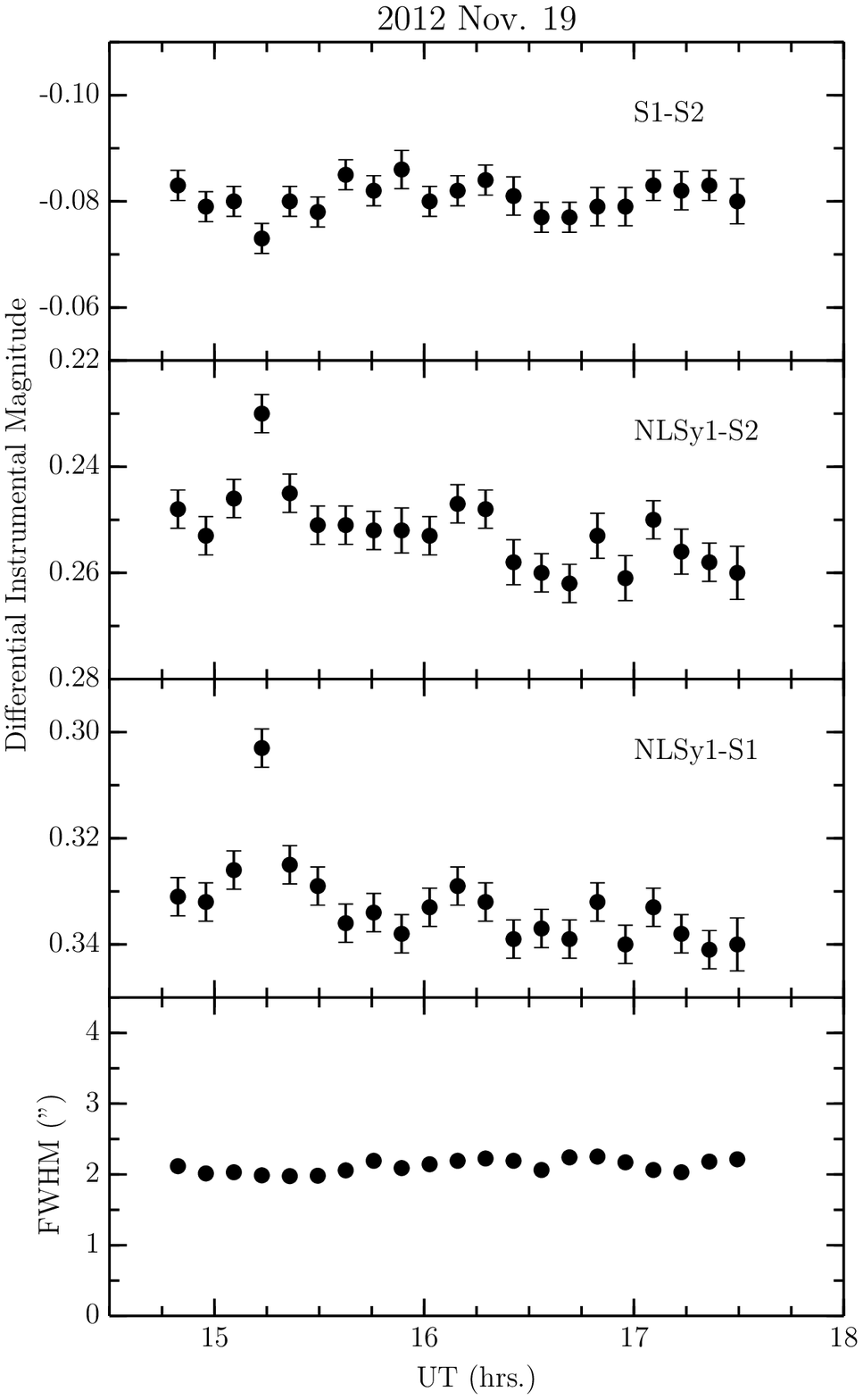}
\includegraphics[width=6.0cm,height=12.0cm]{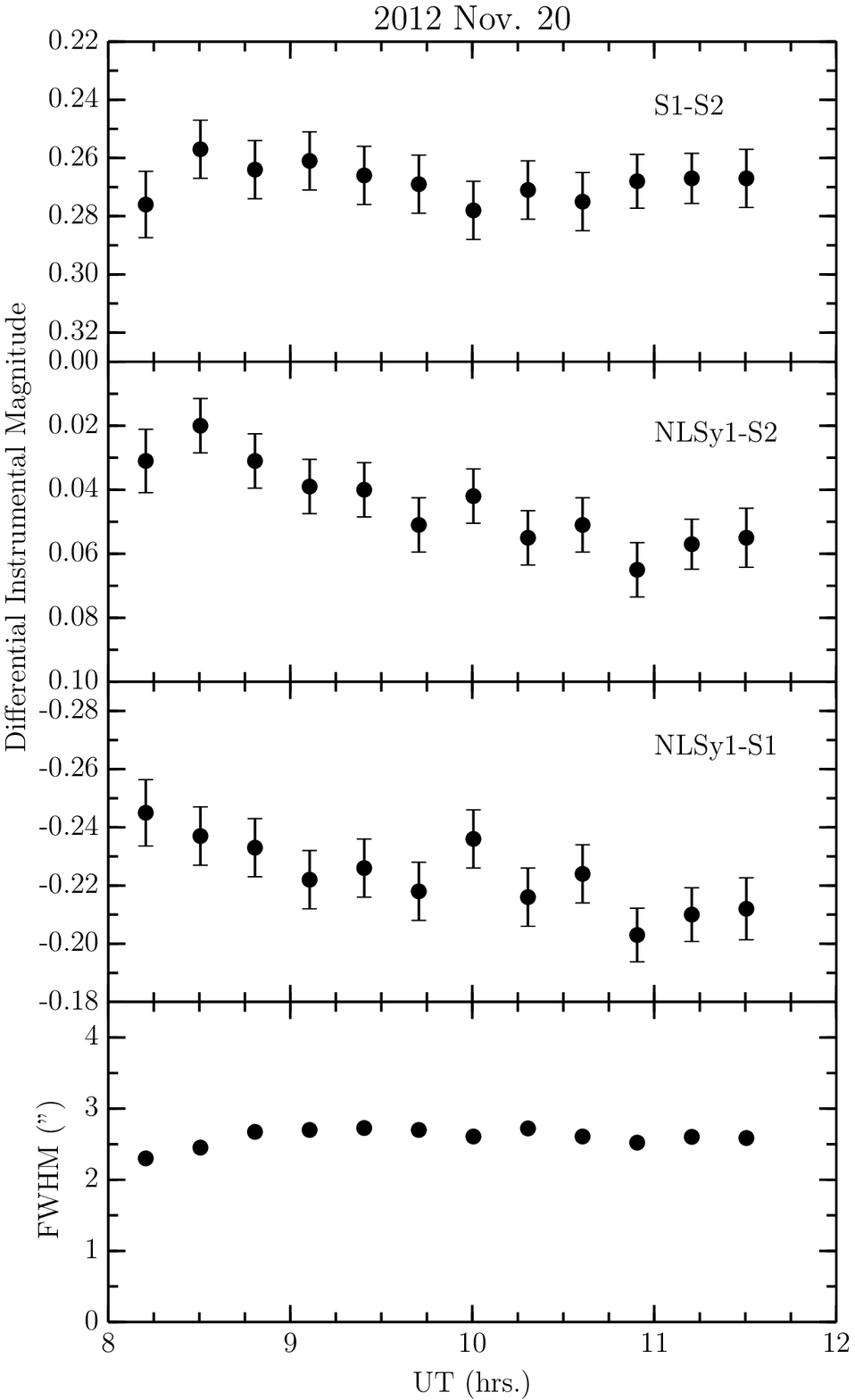}
\includegraphics[width=6.0cm,height=12.0cm]{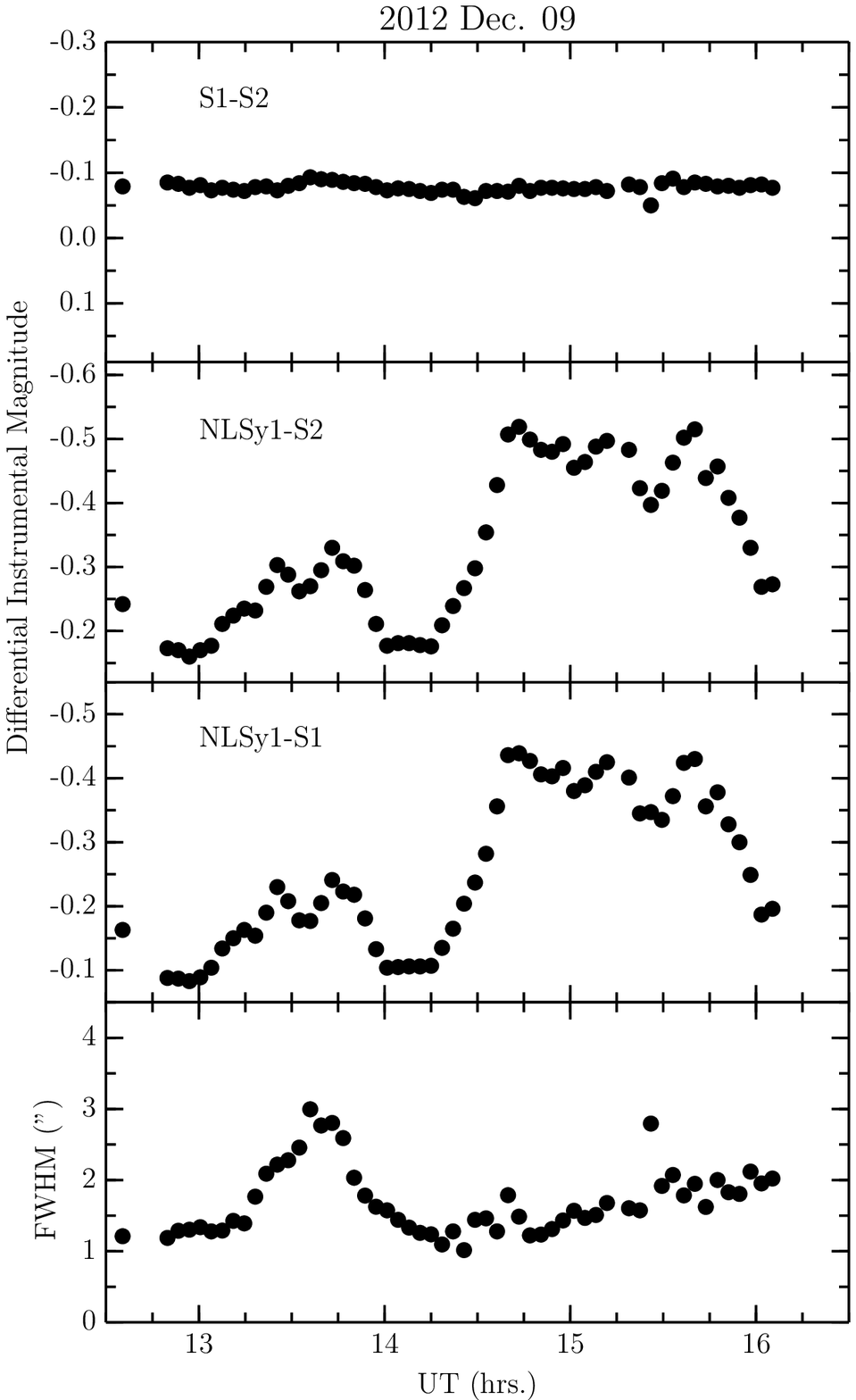}
}
\caption{Intra-night DLCs of 1H 0323+342. On the bottom panel of each night is given the variations of the FWHM of the stellar images during the monitoring period in the night. NLSy1 refers to the target 1H 0323+342, while S1 and S2 denote the two comparison stars selected to generate the differential light curves.}
\label{fig7}
\end{figure*}

We observed 1H 0323+342 three times in 2012 November-December, using ground based optical facilities. The intra-night differential light curves (DLCs) are plotted in Figure~\ref{fig7} and the corresponding statistics are given in Table~\ref{tab6}. We consider the source to be variable only when it shows correlated variations in both amplitude and time relative to the selected pairs of comparison stars. We use two statistical tests namely, {\it C}-statistics and {\it F}-statistics, to judge the variability nature of the source. {\it C} parameter is defined as the ratio of the standard deviations of the source and the comparison star DLCs. A source is considered to be variable only if {\it C} $\geq$ 2.576, which corresponds to a 99\% confidence level \citep{1997AJ....114..565J}. Recently, use of {\it C}-statistics is questioned by \citet{2010AJ....139.1269D}. According to them, a better test to assess the variance in AGN light curve, is the {\it F}-statistics. This method takes into account the ratio of two variances given as {\it F} = $\sigma^{2}_{T}$/$\sigma^{2}_{cs}$, where $\sigma^{2}_{T}$ is the variance of source-comparison star DLC and $\sigma^{2}_{cs}$ is the variance of the comparison star-star DLC. The calculated {\it F} values are then compared with critical {\it F} value, {\it F$^{\alpha}_{\nu}$}, where $\alpha$ is the significance level and $\nu$ (= N$_{p}$ $-$ 1) is the degree of freedom for the DLC. A significance level of $\alpha$ = 0.01 corresponds to a confidence level $>$ 99\%. We assume the source to be variable only if both the computed {\it F} values, corresponding to the DLCs of the source to each of the two comparison stars, are above the critical F{\it } value corresponding to a confidence level $>$ 0.99. We note here that {\it C}-statistics might be a more realistic measure of presence of variability, particularly when the comparison star light curves are not steady.
Further, in the observed CCD frames, variations in the FWHM of the point source might give rise to fictitious variations in the source. However, since we have not found any correlation of FWHM with the variability pattern observed in the DLCs, the observed variability is the genuine variations of the source. 

One of the INOV observations reported here (2012 December 9) coincides with one of the $\gamma$-ray flares. In Figure~\ref{fig1}, this is shown by red downward arrow. During this night, the brightness of the source increased by $\sim$ 0.4 magnitude within 30 minutes (see third panel of Figure~\ref{fig7}, UT 14.25 to 14.75). On top of the large flare, some small but ultra-fast variations ($\sim$ 10-15 minutes) are also visible. Similar fast variability features are already reported by \citet{2013MNRAS.428.2450P} and \citet{2013ApJ...775L..26I} for another $\gamma$-NLSy1 galaxy PMN J0948+0022. This indicates the existence of high incidence of INOV during $\gamma$-ray flaring activity.

\section{Discussion}\label{sec4}
Detection of variable $\gamma$-ray emission from some RL-NLSy1 galaxies, similar to that seen in blazars, confirms the presence of relativistic jets in them. Further, NLSy1 galaxies are thought to be hosted by spirals \citep{2003AJ....126.1690C} with low mass black hole, whereas blazars are hosted by massive elliptical galaxies \citep{2009arXiv0909.2576M}. Therefore, confirmation of the presence of relativistic jets in NLSy1 galaxies has challenged the paradigm that jets can only be hosted by elliptical galaxies. Detailed study of $\gamma$-NLSy1 galaxies are therefore needed to better understand their nature.

\begin{figure*}
\centering
\includegraphics[width=10.0cm]{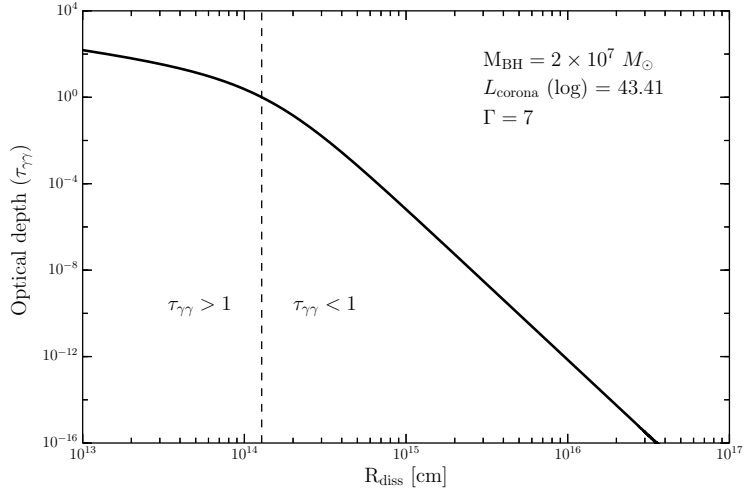}
\caption{Variation of optical depth, for the interaction of 1$-$10~GeV photons with X-ray coronal photons, with distance of emission region from central black hole. Vertical dashed line separate the optically thick and thin region for $\gamma$-ray production.}
\label{fig11}
\end{figure*}

Analysis of the multi-wavelength variability characteristics of 1H 0323+342 can provide hints on 
the jet environment surrounding the emission region. During the period MJD 55000$-$56000, though the source displays flux variations in the optical and X-ray bands, no such significant variability is seen in the $\gamma$-ray band (see Figure~\ref{fig1}). A possible explanation of such behavior could be that the emission region is located close to the central black hole. In such a scenario, first, the observed X-ray spectrum would be soft \citep{2009MNRAS.397..985G} and second, because of $\gamma-\gamma$ interaction with coronal photons, most of the $\gamma$-rays would be absorbed and thus there would be modest $\gamma$-ray emission. Fitting of X-ray spectra by a PL$_X$ model (in most of the observations, except the X-ray flaring states where the BPL model gives a better fit) indeed shows a soft X-ray spectrum. Also, a flare in optical/X-ray band with no counterpart in $\gamma$-rays 
suggests absorption of $\gamma$-ray photons. This could explain uncorrelated variability seen in the light curves. The one zone leptonic model used in the paper cannot explain the SED of this anomalous variability period and a separate study of this peculiar feature will be presented elsewhere. During the later part of the light curve, the $\gamma$-ray flaring activities shown by the source along with the output parameters of SED modeling indicate that the emission region is located far from central black hole, but still inside the BLR. In such a situation, the surrounding jet environment would be transparent to $\gamma$-rays, and thus any flaring activity in other wavelengths will be accompanied by a $\gamma$-ray flaring event. In order to describe this interpretation, we estimated the optical 
depth ($\tau_{\gamma\gamma}$) at different locations of the emission region ($R_{\rm diss}$), using the prescriptions of \citet{1995MNRAS.273..583D} and \citet{2009MNRAS.397..985G}. This calculation is performed for the interaction of $\gamma$-ray photons having energy 1$-$10~GeV with the 
soft X-ray coronal photons. The variation of $\tau_{\gamma\gamma}$ with $R_{\rm diss}$ is shown in Figure~\ref{fig11}.  
It is evident from this plot that most of the $\gamma$-ray emission gets absorbed at $R_{\rm diss}~\lesssim 10^{14}$~cm, indicating absorption as a possible explanation for the uncorrelated variability, if the emitting region is located at $R_{\rm diss} < 10^{14}$ cm during the period MJD 55000$-$56000.

\begin{figure*}
\centering
\includegraphics[width=10.0cm]{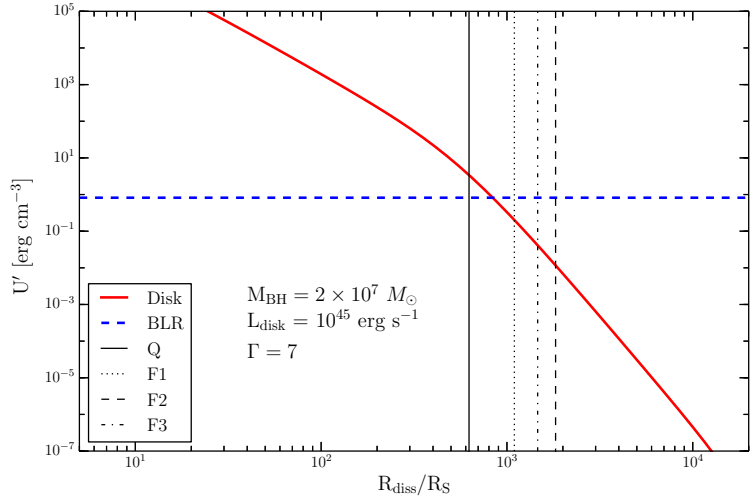}
\caption{Variation of disk and BLR energy densities, measured in the comoving frame, as a function of dissipation distance (in units of Schwarzschild radius). Vertical lines show the location of emission region during the different activity states.}
\label{fig12}
\end{figure*}

From SED modeling of the four different $\gamma$-ray activity states, we find that the optical/UV part is dominated by accretion disk emission. Also, during
flaring states, increase in SSC emission is noticed, which is attributed to increase in magnetic energy density ($U_B$). Under the assumption of equipartition 
between $U_B$ and particle energy density ($U_e$), increase in $U_B$ leads to enhancement of $U_e$ (see Table~\ref{tab5}). The increase of $U_e$ explains the hardening of the 
$\gamma$-ray spectrum as well as an increase of the EC-BLR flux.
Moreover, in the quiescent state SED, EC-disk is found to be dominating over EC-BLR, while in all the three flaring states, 
the EC-BLR peak is higher than EC-disk. The relative dominance of EC-BLR and EC-disk depends upon the relative energy densities of BLR ($U^{\prime}_{\rm BLR}$) and disk 
radiation ($U^{\prime}_{\rm disk}$), measured 
in the comoving frame. These energy densities in turn depend upon the bulk Lorentz factor and the location of the emission region from the central 
source \citep[see][for a detailed discussion]{2009MNRAS.397..985G}. The variation of $U^{\prime}_{\rm disk}$ and $U^{\prime}_{\rm BLR}$ as a function of the distance of the emission region from black hole, for $\Gamma = 7$, is shown in Figure~\ref{fig12}. The location of the emission region during different activity states are demarcated 
as vertical lines\footnote{A negligible deviation in energy densities can be noted for $\Gamma = 8$, bulk Lorentz factor considered for the SED modeling of state F3.}. It is clear from this figure that during the quiescent state $U^{\prime}_{\rm disk}$ is higher than $U^{\prime}_{\rm BLR}$, whereas, in all
the three flaring states, $U^{\prime}_{\rm BLR}$ is higher than $U^{\prime}_{\rm disk}$. Therefore, the relative influence of the two  energy densities, as shown in Figure~\ref{fig12}, explains 
the dominance of EC-disk or EC-BLR during the different activity states of the source.
Further, on comparing the SED output parameters with those of a large sample of blazars studied by \citet{2010MNRAS.402..497G}, we notice that parameters 
such as bulk Lorentz factor and jet powers of 1H 0323+342 are quite similar to low luminosity blazars. There are few differences, such as high accretion disk luminosity ($\sim 0.4 L_{\rm Edd}$ for 1H 0323+342 while $\sim 0.1 L_{\rm Edd}$ in blazars) and low black hole mass. In comparison with other $\gamma$-NLSy1 galaxies, studied in their different activity states \citep{2013MNRAS.436..191D,2012A&A...548A.106F}, 1H 0323+342 hosts a relatively weaker jet as compared to SBS 0846+513 and PMN J0948+0022 but has a similar accretion disk luminosity to PMN J0948+0022 which is higher than that of SBS 0846+513. Also, 1H 0323+342 hosts a central black hole with a mass lying intermediate to the high black hole mass $\gamma$-NLSy1 galaxies SBS 0846+513 and PMN J0948+0022 \citep{2008ApJ...685..801Y} and the low black hole mass $\gamma$-NLSy1 galaxies  PKS 1502+036 and PKS 2004$-$447 \citep{2008ApJ...685..801Y,2001ApJ...558..578O}. 
However, it should be noted that the black hole mass is still a poorly estimated quantity \citep[see for e.g.,][]{2013MNRAS.431..210C}.

The recent GeV flare from 1H 0323+342 features a relatively hard photon index (when compared to other states). Together with the large Compton dominance (see state F3 in Figure~\ref{fig6}), this suggests that during this flare, the bulk of the radiative energy is released prominently at 
high energies. The six hour averaged isotropic $\gamma$-ray luminosity during the maximum of the flare is found to be $L_{\gamma, iso} \simeq$~4 $\times$ 10$^{46}$ erg s$^{-1}$. Correspondingly, the total power emitted in the $\gamma$-ray energy band \citep{1997ApJ...484..108S} is therefore $L_{\gamma, em} \simeq$ $L_{\gamma, iso}/2\Gamma^{2} \simeq$~3.1 $\times$ 10$^{44}$ erg s$^{-1}$ (assuming bulk Lorentz factor $\Gamma$ = 8, as found from modeling of SED) which is a considerable fraction of the kinetic jet power ($\sim$~23\%; $P_{j, kin}$ = 1.38 $\times$ 10$^{45}$ erg s$^{-1}$). This implies that within a few hours during this outburst, a good fraction of total kinetic luminosity carried out by jet is converted into
radiation. This sudden outburst is first noticed in the one day averaged $\gamma$-ray light curve and remains distinctly visible even when we opt for smaller time binning. As mentioned earlier, such rapid outburst is not reported for any other $\gamma$-NLSy1 galaxy.

An examination of the averaged X-ray spectrum from \emph{Swift}-XRT has revealed the presence of features commonly seen in radio-quiet (RQ) NLSy1s, namely a soft excess and a broad excess peaking around the energy of the Fe K-$\alpha$ line. We have demonstrated that these features can be well described by a blurred reflection model, in which both the soft excess and iron line feature arise from relativistic smearing of line emission, caused by coronal X-rays hitting the accretion disk, close to the event horizon.
If this interpretation is correct, we can use X-ray observations of 1H~0323+342 to probe the inner accretion disk, where the jet is launched. With the inclination fixed at a low value of ten degrees, as expected for a source with observed jet emission, we find parameters indicative of a rapidly spinning black hole, where the inner disk is strongly illuminated. This supports the suggestion that high spin may be necessary for jet production in AGN \citep[see][and references therein]{Steiner13}. When this model is extrapolated to the higher energies of the \emph{Swift}-BAT, we find that it underpredicts the observed flux at energies above $\sim50$~keV by a factor of 3--4. This is likely to be due to the presence of emission from the jet in the hard X-ray spectrum, which could potentially be used to study the connection between emission from the X-ray corona and the jet, as has been previously suggested \citep[e.g.][]{Markoff05}. Given that the disk must have a low inclination for jetted emission to be clearly detected, it is unlikely that we are viewing the central X-ray source through a wind from the disk, implying that we have a relatively unobscured view of the inner regions of the disk. This supports the reflection interpretation of the spectrum. The energy flow in this source is found to be radiatively inefficient. The accretion disk luminosity is $\sim 10^{45}$ erg s$^{-1}$, the total coronal luminosity (from spectral modeling) is $\sim 10^{43.4}$ erg s$^{-1}$, and the jet kinetic energy is $>10^{45}$ erg s$^{-1}$.  The parameters from spectral modeling suggest a moderately ionized accretion disk, which is thin and dense to less than 2$R_\textrm{G}$. The black hole is accreting at around the Eddington limit and a considerable amount of the disk energy is extracted into the corona and jet.

In our earlier work \citep{2013MNRAS.428.2450P}, we studied the INOV behavior of 1H 0323+342 and found the hints of INOV characteristics with lower amplitude of variability when compared to another $\gamma$-NLSy1 galaxy PMN J0948+0022. A probable reason could be that the source was in a quiescent state during that observational run. Interestingly, this time we are able to observe it in a high $\gamma$-ray flaring state and find violent INOV behavior (see INOV light curve of 2012 December 09 in Figure~\ref{fig7}) similar to that first reported by \citet{2010ApJ...715L.113L} for PMN J0948+0022. We find an amplitude of variability as high as 36\%.  This indicates a high chance of detecting INOV during $\gamma$-ray flaring states of the source. Though the optical part of the SEDs is dominated by accretion disk emission, we 
attribute the observed large amplitude INOV due to increased non-thermal synchrotron radiation from the jet. Had the INOV be due to 
processes in the accretion disk, similar variability  should have been seen on the other nights as well. From SEDs  too, 
we find that the synchrotron emission during the flaring state (F1) is larger than the quiescent state. Therefore, we conclude that
the observed large amplitude INOV during F1 state is due to the jet flare.
Many mini-flares on top of a large flare are detected and one of the few possibilities to explain this rapid variability could be stronger relativistic beaming. This phenomenon is very well observed in blazars having ultra-relativistic jets with high bulk Lorentz factors. Alternatively, an outburst within a small compact region co-spatial with a relatively large region (blob in a blob, see e.g., \citealt{2011A&A...534A..86T}) can also give rise to such fast temporal evolution. We note here that the recent GeV outburst is not resolvable down to GTI scale i.e. at hour scale and also we find the bulk Lorentz factor of the source to be lower compared to powerful blazars, thus the latter scenario could be a more plausible explanation of very fast variability detected from 1H 0323+342.

Few physical properties such as prominent accretion disk emission, requirement of EC mechanism to explain $\gamma$-ray emission along with the presence of significant curvature in the $\gamma$-ray spectrum of 1H 0323+342 indicates its similarity with FSRQs, albeit at low black hole mass end, whereas the X-ray properties of the source are found to be similar to that of conventional RQ-NLSy1 galaxies. These dual characteristics along with the fact that it is the nearest $\gamma$-NLSy1 galaxy with a detected relativistic jet, makes the study of this source of utmost importance. The {\it Swift}-XRT observations (though with poor spectral resolution) have revealed many peculiar features in the X-ray spectrum of the source. Therefore, observations with facilities having better spectral resolution such as {\it XMM-Newton} and {\it NuSTAR} as well as the forthcoming {\it ASTROSAT} 
would be crucial to study the jet launching region in unprecedented detail.

\section{Summary}\label{sec5}

In this paper, we present a detailed multi-wavelength study of the $\gamma$-NLSy1 galaxy 1H 0323+342. We summarize below the main results of our study.

\begin{enumerate}
\item From multi-frequency light curves we find the presence of many uncorrelated flares in optical/X-ray bands with no counterpart in $\gamma$-ray energy range. The uncorrelated flares are found between MJD 55000$-$56000, whereas after this, the flares are likely to be correlated. However, this cannot be statistically claimed owing to sparseness in the data. The presence/absence of correlated $\gamma-$ ray, X-ray/optical flux variations may be caused because of the emission region located at different jet environments where absorption of $\gamma$-rays plays an important role in the detection of uncorrelated variability from the source.

\item During all the four periods considered for SED modeling, the optical/UV part of the spectrum is dominated by emission from the accretion disk. During the quiescent and first flaring state (F1), X-ray coronal radiation dominates the X-ray spectrum. However, in the subsequent flaring states F2 and F3, the contribution of SSC overpowers the coronal X-ray emission. This is consistent with the increase in magnetic field in flaring states, as we find from SED modeling. In all the activity states, the $\gamma$-ray emission is well modeled by EC of disk photons reprocessed by BLR. We observe a remarkable evolution of the X-ray spectra, wherein the emergence of a hard jetted component is evident during the flaring states. 

\item Good photon statistics of the GeV flare shown by 1H 0323+342 enables us to go for finer binning and a flux doubling time scale as small as $\sim$ 3 hours is noticed. Further, the $\gamma$-ray flare is not resolvable even at GTI time scale and thus hints for the outburst to have occurred at an extremely rapid rate. A good fraction of kinetic jet power is found to be radiated in the form of high energy $\gamma$-ray photons during this flare.

\item Examining the mean X-ray spectra from the \emph{Swift}-XRT and BAT has shown that the source looks very similar to other NLSy1 over the energy range from 0.4--50~keV, showing a soft excess and relativistically blurred Fe line. Above 50~keV in the BAT spectrum we find a significant excess, due to the presence of emission from the jet.
Modeling the reflection spectrum returns a high spin value, $0.96\pm0.14$, and a steep emissivity profile  indicating emission from the very inner regions of the accretion disk. It appears that the inner disk in this jetted source is average, intact, and behaving like other NLSy1s. There should be $2\Gamma^2$ or $\sim100$ times as many similar objects beamed out of our line of sight, which suggests that most NLSy1 should host a jet.

\item The relation of the INOV behavior of a source to its apparent brightness state is not well known. For 1H 0323+342, we find large amplitude ($\sim$~36\%) INOV behavior in its $\gamma$-ray flaring state. This observation clearly hints of the prevalence of high incidence of INOV during the active state of the source.
\end{enumerate}

\acknowledgments
We sincerely thank the referee for constructive comments which helped to improve the manuscript.
This research has made use of data from the OVRO 40-m monitoring program \citep{2011ApJS..194...29R} which is supported in part by NASA grants NNX08AW31G and NNX11A043G, and NSF grants AST-0808050 and AST-1109911. This research has made use of the data obtained from HEASARC provided by the NASA's Goddard Space Flight Center. Part of this work is based on archival data, software or on-line services provided by the ASI Science Data Center (ASDC). This research has made use of the XRT Data Analysis Software (XRTDAS) developed under the responsibility of the ASDC, Italy. The CSS survey is funded by the National Aeronautics and Space Administration under Grant No. NNG05GF22G issued through the Science
Mission Directorate Near-Earth Objects Observations Program. The CRTS survey is supported by the U.S. National Science Foundation under grants AST-0909182 and AST-1313422. This research has made use of the Palermo BAT Catalogue and database operated at INAF $-$ IASF Palermo. Use of {\it Hydra} cluster at Indian Institute of Astrophysics is acknowledged.


{\it Facilities:} \facility{Fermi}, \facility{Swift}, \facility{HCT}.

\bibliographystyle{apj}
\bibliography{ms}

\clearpage
\begin{table}
\begin{center}
{
\small
\caption{General parameters and results of the analysis of about five years of {\it Fermi}-LAT data of 1H 0323+342. Column 
information are as follows: (1) Name; (2) right ascension; (3) declination; (4) redshift; (5) 0.1$-$100 GeV flux in units of 10$^{-8}$ ph cm$^{-2}$ s$^{-1}$; (6) log parabolic photon index at pivot energy obtained from {\it Fermi}-LAT data analysis; (7) curvature index; (8 )TS; (9) $\gamma$-ray luminosity in units of 10$^{44}$ erg s$^{-1}$ and (10) $N_H$ \label{tab1}}
\begin{tabular}{lccccccccc}
\tableline\tableline
Name  & RA (2000) & Dec (2000) & z\tablenotemark{a} & F$_{0.1 - 100 GeV}$ &  $\alpha$ & $\beta$ & TS & L$_{\gamma}$& $N_H$\tablenotemark{b}\\
      &  (h m s)  & (d m s)    &                    &                     &           &         &    &            &                     \\
(1)   & (2)       & (3)        & (4)                &  (5)                &  (6)                & (7)& (8) & (9)   & (10)          \\
\tableline
1H 0323+342   & 03:24:41.2 & +34:10:45 & 0.063 & 6.80 $\pm$ 0.42 & 2.67 $\pm$ 0.07 & 0.41 $\pm$ 0.09 & 621.88 & 2.92 & 1.27 \\

\tableline
\end{tabular}
\tablenotetext{1}{\citet{2010A&A...518A..10V}}
\tablenotetext{2}{Galactic absorption in units of 10$^{21}$ cm$^{-2}$ from \citet{2005A&A...440..775K}}
}
\end{center}
\end{table}

\begin{table}
\begin{center}
{
\small
\caption{Fractional root mean square variability amplitude ({\it F$_{var}$}) values for different energy bands, calculated for the light curves shown in Figure~\ref{fig1}.\label{tab2}}
\begin{tabular}{cc}
\tableline\tableline
Energy band & {\it F$_{var}$} \\
\tableline
Radio (15 GHz)  & 0.351~$\pm$~0.003 \\
V (UVOT)        & 0.062~$\pm$~0.005\\
B               & 0.098~$\pm$~0.005\\
U               & 0.118~$\pm$~0.006\\
UVW1            & 0.125~$\pm$~0.007\\
UVM2            & 0.141~$\pm$~0.007\\
UVW2            & 0.130~$\pm$~0.006\\
X-ray (0.3-10 keV)& 0.340~$\pm$~0.004\\
X-ray (15-150 keV)& 0.000~$\pm$~0.000\\
$\gamma$-ray (0.1-100 GeV) & 0.404~$\pm$~0.056\\
\tableline
\end{tabular}
}
\end{center}
\end{table}

\begin{table}
\begin{center}
{
\small
\caption{Results of model fits to the $\gamma$-ray spectra averaged over different activity states considered in Figure \ref{fig1}.\label{tab3}}
\begin{tabular}{ccccccccc}
\tableline\tableline
 &  & Power-Law &  &  &  &  & &\\
\tableline
Activity state & Period & Flux$_{0.1-100 GeV}$              & Luminosity              & $\Gamma_{\gamma}$    & TS  & {\it N$_{pred}$}\tablenotemark{a} & &\\
               & (MJD)  & (10$^{-7}$ ph cm$^{-2}$ s$^{-1}$) & (10$^{44}$ erg s$^{-1}$)&             &     &  & &\\
\hline
 P      & 54683$-$56200 & 0.57 $\pm$ 0.05 & 1.94 & 2.85 $\pm$ 0.07 & 260.75  & 2806.14 &  & \\
A1      & 56200$-$56400 & 1.46 $\pm$ 0.06 & 5.32 & 2.74 $\pm$ 0.04 & 201.78  & 881.10  &  & \\
A2      & 56400$-$56550 & 1.90 $\pm$ 0.16 & 7.34 & 2.66 $\pm$ 0.08 & 291.19  & 932.90  &  & \\
\hline
 &   & Log-parabola &  &  &  &  & &\\
\hline
Activity state  & Flux$_{0.1-100 GeV}$      & Luminosity   &$\alpha$  & $\beta$ & TS  & {\it N$_{pred}$}\tablenotemark{a} & 2$\Delta${\it L}\tablenotemark{b} &\\ 
                & (10$^{-7}$ ph cm$^{-2}$ s$^{-1}$) & (10$^{44}$ erg s$^{-1}$)&             &         &     &  & &\\
\hline
P       & 0.50  $\pm$ 0.05 & 2.20  & 2.78 $\pm$ 0.12 & 0.62 $\pm$ 0.17  & 284.32  & 2715.66 & 30.5 &\\
A1      & 1.29  $\pm$ 0.09 & 5.56  & 2.59 $\pm$ 0.06 & 0.30 $\pm$ 0.05  & 206.21  & 840.64 & 9.6  &\\
A2      & 1.81  $\pm$ 0.16 & 7.58  & 2.58 $\pm$ 0.09 & 0.16 $\pm$ 0.08  & 295.80  & 920.69 & 4.9  &\\
\hline

\tableline
\end{tabular}
\tablenotetext{1} {Number of predicted photons detected in the time period considered}
\tablenotetext{2}{$\Delta${\it L} represents the difference of the logarithm of the likelihood with respect to a single power-law fit }
}
\end{center}
\end{table}

\begin{table}
\begin{center}
{
\small
\caption{Results of reflection model fits to the XRT average spectrum, shown in Figure~\ref{plratio}. Columns are as follows: (1) Column density to the source in cm$^{-2}$; (2) Photon index of the primary power-law; (3) Black hole spin; (4) and (5) inner and outer emissivity indices, respectively; (6) Break radius of the emissivity profile, in gravitational radii;(7) ionization of the reflection component, in erg s$^{-1}$ cm$^{-2}$; (8) Inclination in degrees. \label{xrayfittable}}
\begin{tabular}{cccccccc}
\tableline\tableline
$N_\textrm{H}$		&	$\Gamma_{X}$		& a	&$q_1$  	&	$q_2$&	$R_\textrm{break}$\tablenotemark{a}	&	$\xi$ 	&	$\theta$\tablenotemark{a} \\
(1)	&	(2)	&(3)&(4)&(5)&(6)&(7)&(8)\\
\tableline
$1.27\times 10^{21}$&	$2.02\pm0.06$	&$0.96\pm0.14$&	$>7.5$	&	$1.8\pm0.4$&6	&	$212\pm12$& 10\\
\tableline
\end{tabular}
\tablenotetext{1} {The break radius of the emissivity profile and the inclination of the disk are fixed at these values.}
}
\end{center}
\end{table}

\begin{table*}
\begin{center}
{
\scriptsize
\caption{Summary of SED analysis\label{tab4}}
\begin{tabular}{ccccccccc}
\tableline\tableline
 & & & {\it Fermi}-LAT &  & & & &  \\
\tableline
 Activity state & Period\tablenotemark{a} & Flux\tablenotemark{b} & Photon Index\tablenotemark{c} & Test Statistic\tablenotemark{d} & {\it N$_{pred}$}\tablenotemark{e}& & &  \\
 \tableline
 Q  & 54775$-$54805  &1.36~$\pm$~0.28 & 2.83~$\pm$~0.20 & 40.90  & 198.65 & & &  \\
 F1 & 56262$-$56310  &2.54~$\pm$~0.28 & 2.56~$\pm$~0.10 & 184.19 & 431.40 & & &  \\
 F2 & 56470$-$56500  &3.57~$\pm$~0.38 & 2.63~$\pm$~0.10 & 225.05 & 401.50 & & &  \\
 F3 & 56531$-$56535  &9.95~$\pm$~1.43 & 2.43~$\pm$~0.14 & 157.04 & 123.67 & & &  \\
 \tableline
 & &  & {\it Swift}-XRT & & &  & &\\
 \tableline
 Activity state & Exp.\tablenotemark{f}& $\Gamma_{1}$\tablenotemark{g}& $\Gamma_{2}$\tablenotemark{h} & Flux\tablenotemark{i} & Normalization\tablenotemark{j}& Stat.\tablenotemark{k} &  &\\
\tableline
 Q  & 5.94  & 1.94~$\pm$~0.06 &               & 11.41$^{+0.87}_{-0.84}$ & 2.56~$\pm$~0.10 & 62.59/49   & &\\
 F1 & 10.09 & 1.95~$\pm$~0.03 &               & 11.84$^{+0.50}_{-0.49}$ & 2.67~$\pm$~0.06 & 117.82/118 & &\\
 F2 & 16.89 & 2.05~$\pm$~0.04 & 1.58$\pm$0.09 & 23.89$^{+0.89}_{-0.87}$ & 4.82~$\pm$~0.08 & 244.47/229 & &\\
 F3 & 1.98  & 1.66~$\pm$~0.06 &               & 29.54$^{+2.43}_{-2.36}$ & 4.92~$\pm$~0.23 & 42.11/40   & &\\
 \tableline
 & & & {\it Swift}-UVOT & & & &  &\\
 \tableline
 Activity state & V\tablenotemark{l}& B\tablenotemark{l}& U\tablenotemark{l}& W1\tablenotemark{l}& M2\tablenotemark{l}& W2\tablenotemark{l}& &\\
 \tableline
 Q  & 1.89~$\pm$~0.06& 1.92~$\pm$~0.07& 2.02~$\pm$~0.08& 1.95~$\pm$~0.10& 2.22~$\pm$~0.11& 2.12~$\pm$~0.11& &\\
 F1 & 1.91~$\pm$~0.04& 1.85~$\pm$~0.04& 2.09~$\pm$~0.05& 2.00~$\pm$~0.06& 2.45~$\pm$~0.07& 2.21~$\pm$~0.07& &\\
 F2 & 2.16~$\pm$~0.04& 2.23~$\pm$~0.04& 2.43~$\pm$~0.05& 2.26~$\pm$~0.06& 2.54~$\pm$~0.07& 2.38~$\pm$~0.05& &\\
 F3 & 2.14~$\pm$~0.07& 2.20~$\pm$~0.09& 2.47~$\pm$~0.11& 2.10~$\pm$~0.12& 2.34~$\pm$~0.13& 2.28~$\pm$~0.12& &\\
 \tableline
\end{tabular}
\tablenotetext{1}{Time period considered for SED modeling, in MJD}
\tablenotetext{2}{Integrated $\gamma$-ray flux in 0.1$-$100 GeV energy range in units of 10$^{-7}$ ph cm$^{-2}$ s$^{-1}$.}
\tablenotetext{3}{Photon index calculated from $\gamma$-ray analysis.}
\tablenotetext{4}{Significance of detection using likelihood analysis.}
\tablenotetext{5}{Number of predicted photons during the time period under consideration.}
\tablenotetext{6}{Net exposure in kiloseconds.}
\tablenotetext{7}{Photon index of PL$_{X}$ model or photon index before break energy in BPL model.}
\tablenotetext{8}{Photon index after break energy in BPL model.}
\tablenotetext{9}{Observed flux in units of 10$^{-12}$ erg cm$^{-2}$ s$^{-1}$, in 0.3$-$10 keV energy band.}
\tablenotetext{10}{Normalization at 1 keV in 10$^{-3}$ ph cm$^{-2}$ s$^{-1}$ keV$^{-1}$.}
\tablenotetext{11}{Statistical parameters:$\chi^{2}$/dof.}
\tablenotetext{12}{Average flux in {\it Swift} V, B, U, W1, M2 and W2 bands respectively, in units of 10$^{-11}$ erg cm$^{-2}$ s$^{-1}$.}
}
\end{center}
\end{table*}

\begin{table*}
\begin{center}
{\scriptsize
\caption{Summary of source parameters used to model the SEDs.\label{tab5}}
\begin{tabular}{lcccc}
\tableline
\tableline
Parameter                         &            Q     &         F1          &      F2        &      F3        \\
\tableline
log $L_{d}$ \tablenotemark{a}                       & 45               & 45                   & 45              & 45         \\
log $L_{c}$ \tablenotemark{b}                       & 43.41            & 43.41                & 43.41           & 43.41         \\
{\it p} \tablenotemark{c}                           & 1.2              & 1.2                  & 1.6             & 1.6          \\
{\it q} \tablenotemark{d}                           & 4.9              & 4.2                  & 4.3             & 3.9          \\
{$B_{eq}$}\tablenotemark{e}                         & 7                & 5.3                  & 6.2             & 6.4           \\
$\Gamma$ \tablenotemark{f}                          & 7                & 7                    & 7               & 8             \\
$U_{e}$\tablenotemark{g}                            & 1.95             &  1.12                & 1.53            & 1.63          \\
$\gamma$\textquotesingle$_{min}$  \tablenotemark{h} & 15               & 15                   & 15              & 10             \\
$\gamma$\textquotesingle$_{b}$  \tablenotemark{i}   & 150              & 150                  & 75              & 70           \\
$\gamma$\textquotesingle$_{max}$ \tablenotemark{j}  & 2000             & 2000                 & 2000            & 2000          \\
R$_{BLR}$ \tablenotemark{k}                         & 0.03             & 0.03                 & 0.03            & 0.03          \\
R$_{diss}$ \tablenotemark{l}& 625 (1.2 $\times$ 10$^{-3}$)& 1096 (2.1 $\times$ 10$^{-3}$) & 1827 (3.5 $\times$ 10$^{-3}$)& 1463 (2.8 $\times$ 10$^{-3}$)          \\
\hline
$P_{r}$ \tablenotemark{m}                           & 41.29            & 41.36                & 41.74           & 41.62         \\
$P_{j}$ \tablenotemark{n}                           & 44.06            & 44.22                & 45.03           & 45.14         \\
\tableline
\end{tabular}
\tablenotetext{1}{Accretion disk luminosity in log scale.}
\tablenotetext{2}{Corona luminosity in log scale.}
\tablenotetext{3}{Slope of particle spectral index before break energy.}
\tablenotetext{4}{Slope of particle spectral index after break energy.}
\tablenotetext{5}{Magnetic field in Gauss.}
\tablenotetext{6}{Bulk Lorentz factor.}
\tablenotetext{7}{Particle energy density in erg cm$^{-3}$.}
\tablenotetext{8}{Minimum Lorentz factor of electrons.}
\tablenotetext{9}{Break Lorentz factor of electrons.}
\tablenotetext{10}{Maximum Lorentz factor of electrons.}
\tablenotetext{11}{Size of BLR in parsec.}
\tablenotetext{12}{Distance of the emission region from central black hole in units of Schwarzschild radius and (in parenthesis) in parsec.}
\tablenotetext{13}{Radiative jet power in log scale.}
\tablenotetext{14}{Kinetic jet power in log scale.}
}
\end{center}
\end{table*}

\begin{table*}
\caption{Log of INOV observations. Columns:- (1) date of observation; (2) INOV amplitude in percent; (3)  and (4) {\it F}-values computed for DLCs relative to the steadiest pair of comparison stars on any night; (5) variability status according to {\it F}-statistics, V: variable, NV: non-variable; (6) and (7) Values of {\it C} for the DLCs relative to the two comparison stars and (8) variability status as per {\it C}-statistics.} 
\begin{center}
\begin{tabular}{ccrrcrrc}
\hline \hline
Date     & $\psi$ & {\it F1}  & {\it F2} & Status  & {\it C1}  & {\it C2}  &  Status \\
yy.mm.dd & (percent) &           &          &         &           &           &         \\
 (1)     & (2)    & (3)       & (4)      & (5)     & (6)       & (7)       & (8)     \\
\hline
12.11.19 &  3.41  &  5.30     & 3.69     & V       & 2.73      & 2.37      &  NV     \\
12.11.20 &  4.34  &  4.88     & 3.81     & NV      & 2.23      & 1.77      &  NV     \\
12.12.09 &  35.74 &  279.81   & 276.54   & V       & 16.63     & 16.63     &  V      \\
 \hline
\end{tabular}
\label{tab6}
\end{center}
\end{table*}



\clearpage

\clearpage



\clearpage



\clearpage




\end{document}